# Consumer Research with Projective Techniques: A Mixed Methods-Focused Review and Empirical Reanalysis


**Stephen L. France (Corresponding Author)**

Mississippi State University

Mailstop 9582, Mississippi State, MS 39762

Phone: +1 662-325-1630, Fax: +1-662-325-7012

sfrance@business.msstate.edu




**Consumer Research with Projective Techniques: A Mixed Methods-Focused Review and Empirical Reanalysis**


**Abstract**

This article gives an integrative review of research using projective methods in the consumer research domain. We give a general historical overview of the use of projective methods, both in psychology and in consumer research applications, and discuss the reliability and validity aspects and measurement for projective techniques. We review the literature on projective techniques in the areas of marketing, hospitality & tourism, and consumer & food science, with a mixed methods research focus on the interplay of qualitative and quantitative techniques. We review the use of several quantitative techniques used for structuring and analyzing projective data and run an empirical reanalysis of previously gathered data. We give recommendations for improved rigor and for potential future work involving mixed methods in projective techniques.

**Keywords**: Projective techniques, consumer research, content analysis, mixed methods, correspondence analysis, chi-squared tests




# INTRODUCTION

Projective techniques, adapted from clinical psychology, have a long history in consumer research. Projective techniques are considered to fall under the banner of qualitative research, but often generate order and category information that can be analyzed qualitatively. The area of projective techniques is broad, and projective techniques range from simple word association and sentence completion exercises to complex creative tasks. Projective techniques have been utilized in academic and practitioner consumer research to elicit information on consumer behavior, but do not have the wide usage or visibility of survey techniques. This is partially due to a lack of tools to interpret the results of projective techniques and a lack of metrics to evaluate the quality of the results. The focus of this research is on apprising the current use of projective techniques with a particular emphasis on the use of mixed methods and on giving methodological advice for applying projective techniques in a mixed methods context. There is a strong focus on understanding how projective techniques are used in practice and how the analysis of results from projective techniques can be improved and systematized.

This research aims to fill several gaps in the literature. First, we give a detailed analysis of projective technique-based research in three different subareas of consumer research, marketing, consumer & food science, and hospitality & tourism. Research streams in these subdisciplines have developed in parallel, with little methodological overlap, particularly between the more qualitative-focused research found in marketing and hospitality & tourism, and the more quantitative-focused research found in consumer & food science. We compare and contrast the use of projective techniques in these areas and apprise the use of quantitative methods for analyzing results. Second, where data are available, we perform an empirical re-analysis of results derived from projective techniques, focusing on the widespread use of association tests



and correspondence analysis. We use this analysis, combined with work on best practices in methodology, to give advice on analyzing data from projective techniques in a mixed methods environment. Third, we apprise the current status of projective methods in consumer research in light of new methodological developments, such as increased use of quantitative methods for the analysis of unstructured, qualitative data, and give advice for future development of methodology for projective techniques.

## THE CONSUMER RESEARCH ENVIRONMENT

Consumer researchers need to gather information to help understand consumer behavior and perceptions. From a strategy perspective, having a deep understanding of consumer wants and needs is critical to appropriately target and implement effective consumer marketing strategies (Kotler & Keller, 2016), to personalize marketing offerings (Vesanen & Raulas, 2006), and to maximize consumer value (Holbrook, 2006). Despite changes in the marketing environment brought about by the global Covid-19 pandemic, and by digital marketing transformation and the growth of AI, the need for marketing research to understand customers is still paramount. A 2023 survey from Qualtrics (Qualtrics, 2023) found that despite an overall decrease in marketing budgets, marketing research budgets have continued to increase and that there is a need for marketing researchers who have the skills to turn data into marketing insights and effectively utilize new AI technologies.

While the explosion of available marketing data and new data sources, particularly from e-commerce and cloud, has brought about a renewed focus on data analytic methods in marketing research (e.g., Iacobucci et al., 2019, Wedel & Kannan, 2016), traditional marketing research, including survey research, and in-depth qualitative research, still remain important. For example, the online survey software market is worth $5 billion per year, and is expected to grow at a



compound rate of over 11% over the next seven years. Quantitative survey instruments can provide measurements of consumer views and purported consumer behavior. However, there is often a gap between consumers' self-reported behavior from surveys and actual behavior. To help fill this gap, qualitative research is often used to gain additional insight and develop a deeper, more nuanced understanding of consumer behavior (e.g., Price et al., 2015; Fischer & Guzel, 2023). While much traditional qualitative research, carried out using methods such as focus groups, in-depth interviews, and observational ethnographic research, has been done in person, with small sample sizes, digital transformation has created a range of opportunities for online, digital qualitative research. Frameworks and tools have been developed for qualitative analysis in an online setting. For example, netnography (Kozinets, 2002) gives a framework for ethnographic analysis in virtual communities, and focus groups (e.g., Stewart & Shamdasani, 2017) and in-depth interviews (e.g., Keen et al., 2022) are increasingly performed online. In addition, machine learning and AI tools have to been used to help supplement the labor intensive process of analyzing quantitative data for large online datasets, such as social media messages and online reviews (e.g., Berger et al., 2020) and for automating traditional qualitative procedures, such as coding (Zambrano et al., 2023) and content analysis (e.g., Boumans & Trilling, 2018). Likewise, in experimental research, traditional methods such as advertising viewing responses and taste testing have been supplemented with a range of new technologies including virtual reality (e.g., Yoon & Zou, 2024) and biometric testing (e.g., Harris et al., 2018).

In the environment outlined above, projective techniques fill an interesting niche. When implemented well they can help researchers gain deeper understanding of consumers than can be gained from purely quantitative methods. However, these techniques have a "niche" positioning, are often poorly understood, and do not have the evaluative metrics and mechanisms available



for quantitative research and more structured types of qualitative methods, such as content analysis. There is scope for improving this understanding, particularly given the current advances, where machine learning and AI methods are being used to augment and improve qualitative analysis. This is the focus of the remainder of the paper. We first give an overview of projective techniques and the history of projective techniques in consumer research. We describe and categorize the different types of projective techniques and discuss reliability and validity concerns encountered with projective techniques. We give a detailed, systematic review of the literature of projective techniques in three main areas of consumer research, marketing, hospitality & tourism, and consumer & food science, run an empirical reanalysis of datasets found during the systematic review, and use these analyses to help give recommendations for the analysis of projective technique data and for future work in developing projective technique-based methodology.

## PROJECTIVE TECHNIQUES

Projective techniques have a long history in clinical psychology and psychoanalysis, and are utilized in situations where clinicians do not wish to ask sensitive questions directly or where subjects are "repressing" certain uncomfortable memories (e.g., Mosak, 1958). The overall rationale behind projective techniques is that "individuals' personality characteristics, needs, and life experiences" (Lilienfeld et al., 2000) are influenced by how they interpret and respond to what are often classed as "ambiguous" stimuli. Unstructured or ambiguous stimuli will give a higher level of projection of "emotions, motives, attitudes and values" (Donoghue, 2000). A stimuli may not necessarily make sense on its own, but by filling in the details for a given scenario or task, the subject "projects" their own mental processes and motivations onto the task. For example, in the classic Rorschach ink blot test, subjects are asked to describe and interpret



the meaning of multiple "ink-blot images. Though there are no specific correct interpretations, the way the subjects approach the task and the meanings that they attach to the images can be categorized and coded to give insights into the subjects' mental processes (e.g., Mihura et al., 2013).

In marketing, projective techniques have long been used to help gain a deeper understanding of consumer wants, needs, and motivations. Several academic marketing articles in the 1950s and 1960s attempted to apprise the use of projective techniques and offered recommendations for their use. Haire (1950) gave an example where subjects were shown two shopping lists, identical but for the brand of the coffee, with one instant brand of coffee (Nescafe) and one traditional brand (Maxwell House). By asking subjects to project a hypothetical shopper's "personality and character" from one of the shopping lists, managers could gain insight into perceptions of the products or product categories (in this case, instant coffee) and use this to alter marketing strategy. For example, in the Maxwell House case, advertising could help mitigate the perception of the "laziness" of instant coffee users. Zober (1956), described the use of the PF (Picture Frustration) and TAT (thematic apperception test) methods. In the PF method, respondents are presented with an image and conversations between characters (often in conversation bubbles) and then need to complete the sentences. For example, picturing a conversation between two shoppers, one person might say "I like shopping at store X as the most fashionable people shop there" and the second person will respond "I like shopping…..", where the remainder of the sentence is to be filled in by the subject, who will have free range to mention the specific store or some other aspect of shopping. Similarly, in the TAT method, respondents are shown stimuli and are asked to create a story related to the stimuli, for example, an image of a fashion retailer with a large "For Sale" sign.



Compared with traditional survey research methods, projective techniques have some specific advantages when eliciting consumer behavior. By utilizing ambiguous or indirect stimuli, projective techniques are seen as less intrusive (Pich et al., 2015), are less prone to social desirability bias (Fisher, 1993; Will et al. 1996), can help break down cultural barriers (e.g., Broddy, 2004; Dana, 2005; Landgarten, 2017; Pich et al., 2015; Ramos, 2007), and can help shed light on consumer decisions that are not entirely rational, the so-called concept of "bounded rationality" (Bond & Ramsey, 2010)

Since the initial marketing applications of projective techniques in the 1950s, projective techniques have become an accepted part of marketing research literature. An excellent review by Campos et al. (2020) provides an overview of the current status and topics of interest for projective techniques in marketing. It is apparent that projective techniques have found favor in certain niche areas adjacent to marketing, such as hospitality & tourism, and consumer & food science. However, the use of projective techniques by mainstream academic marketing researchers is limited, with a high proportion of the published work appearing in a few qualitative research- and practitioner- focused journals (Campos et al., 2020). Potential reasons for this include the complexity of projective techniques and the need for highly trained researchers (Donoghue, 2000), the difficulty of evaluating projective techniques, and the perceived lack of rigor and lack of publishing opportunities for qualitative work in top journals (e.g., Levy, 2006). However, the major focus of this article is not a rumination on the state of projective methods in consumer research, but the evaluation of methods and metrics for analyzing the results and improving the methodological rigor of projective techniques. A necessary condition for this task is to understand the scope of projective techniques, including the different sub-categories of projective techniques and the commonalities and differences



between techniques in these categories in the context of consumer research. Accordingly, we give an analysis of projective technique categories below.

## Categorization of Projective Techniques

Projective techniques can range from simple structured word selection from a list of pre-defined words to complex creative endeavors, where subjects are asked to create representations such as stories, pictures, and models. There are several methods of categorizing projective techniques. The most common categorization is given by Lindzey (1959) and adapted by numerous authors (e.g., Bond & Ramsey, 2010; Lilienfeld et al., 2000; Pich et al., 2015). Here, techniques are categorized into ordering and arrangement, association, completion, construction, and creative/expression techniques.

Consider a simple branding example where the researcher is looking to gain insight into how users feel about a certain retail brand. An ordering and arrangement task would be the most structured. For example, a respondent could be shown pictures of people and be asked "Order these people by the likelihood of shopping at this retailer". Alternatively, they could be given a list of "adjectives", describing the store environment, e.g., "warm", "happy", and "welcoming", and then rank the adjectives in order of applicability. In an association task, respondents could be asked to the give the celebrity, animal, or car that they would associate with the brand, and in addition, give further associations for the associated item (e.g., adjectives describing the animal). In a completion task, users could be given a PF (Picture Frustration) stimuli with two shoppers, with the first saying "I shop at Store X at least once a week" and the second saying "I also shop here, because…….", with the second sentence needing completion. Construction could range from story creation, as in the previous TAT example to more complex tasks, such as the construction of a model "idealized" retail space. Expression could involve tasks as varied as



creating a sketch or play involving some aspect of the retail experience to creating pictures or videos.

There is no firm boundary between the different categories. For example, a free-form association task of "celebrities" could be reduced to an ordering and arrangement task if the subject is asked to give the celebrities from an ordered list. A key aspect here is the level of structure, both in the projective test and in the data gathered from the projective test. An ordering projective test will produce "pre-structured' data that can easily be analyzed. Other more creative tasks, with more abstract stimuli, may require coding before any summarization and processing. Another consideration is the level of abstractness of the stimuli. How close are the projections to the actual variable or construct of interest? There is a trade-off in the level of "abstractness" and relation to the objects of interest. As noted previously, more unstructured and ambiguous tasks may provide more projective insights (Donoghue, 2000). However, if the tasks are too abstract, they may lack validity with respect to the problem domain (Lilienfeld et al, 2000).

**Evaluating Projective Techniques**

When gathering and analyzing the consumer data, researchers need to ensure that the data collection and analysis process is both "reliable" and "valid". The concepts of reliability and validity are core to effective, replicable research in the behavioral sciences. Reliability is concerned with the replicability of results. For example, consider a scenario where a researcher wishes to measure a consumer's "life satisfaction". If the measurement instrument and experimental instructions do not change, there should be similar results across multiple tests, giving so-called test-retest reliability. Similarly, if multiple trained raters are asked to categorize a consumer's "brand attachment" based on qualitative social media posts or blog entries, the categorization should show a high degree of consistency across raters. Such consistency can be



analyzed through qualitative-based procedures for building consensus or through quantitative measures of agreement, such as Cohen's kappa (McHugh, 2012) or Krippendorff's alpha (Hayes & Krippendorff, 2007). Similarly in survey research, when measuring a factor or construct with multiple items, these items should show a degree of reliability, with high correlations between the items, calculated using a measure such as Cronbach's alpha (Chronbach, 1951), and items loading onto latent constructs using exploratory factory analysis (e.g., Spector, 1992). For example, a two-factor measure of brand attachment (Park et al., 2010) contains two factors, "Brand–Self Connection" and "Prominence", each containing highly correlated sub-items.

Validity is harder to conceptualize and measure than reliability. External validity can be thought of as a measure of whether a measure or construct can be extrapolated to and is valid in a population (Lucas, 2003). Sampling error and non-response error or bias are major threats to external validity (France et al., 2024). In an experimental context, internal validity can be conceptualized in terms of the validity between the variables in an experiment (Cahit, 2015), and threats to internal validity can come from confounding variables and unobserved covariates. Construct validity (Peter, 1981) is the most commonly measured type of validity in consumer research. A construct showing a high level of construct validity should be positively correlated to hypothesized similar scales (convergent validity) and negatively correlated to hypothesized opposite scales (divergent validity). For example, a brand attachment scale should be positively correlated to similar scales of brand attachment, and a scale measuring brand love should be negatively correlated to a scale of brand dislike or hate.

In the context of projective techniques, one of the major critiques of the traditional applications of projective techniques in clinical psychology, for example, the Rorschach ink blot test, has been a lack of strong evidence of reliability and validity (Lilienfeld, 1999). However, with



increased emphasis on standardization and more detailed implementation guides, for example Comprehensive System (CS) for the Rorschach test (Exner Jr, & Erdberg, 2005), there have been a large number of studies showing satisfactory reliability (e.g., Hibbard et al. 2001; Meyer et al. 2012; Viglione et al., 2012) and validity (Lundy, 1988; Mihura et al., 2013; Weiner, 1996) for properly administered projective techniques. A detailed synthesis of the extent literature as of the turn of the millennium is given in Lilienfeld (2000). There are recent summaries of empirical reliability and validity for tests including the Rorschach test, with a focus on showing that these tests meet the required standards for legal evidence (de Ruiter et al., 2023; Viglione et al., 2022). McGrath et al. (2023) weighed recent evidence for the three most commonly used projective tests, the Rorschach ink blot test, the TAT, and figure drawings, and found evidence for reliability and validity, but disputed the use of the nomenclature of "projective" and "test" for these techniques.

In the clinical psychology setting outlined above, rigor is defined in terms of well-known quantitative reliability metrics. This is due to the general standards and quantitative orientation of the medical field, and also the potential to do harm with bad diagnoses. For example, the Rorschach test is often used to diagnose psychopathy (Wood et al., 2010). In a consumer research setting, quality metrics are much less standardized and well-defined. However, the reliability and validity of research instruments is still important in a consumer research setting and quantitative metrics for reliability and validity are widely used in some forms of qualitative research, particularly in content analysis, which is something of a "boundary" method between qualitative and quantitative research methods, with advocates of purer qualitative analysis (e.g., Kuckartz & Rädiker, 2023), and of incorporating quantitative metrics into content analysis (e.g., Krippendorff, 2004; Oleinik et al., 2014).



It may be that small, convenience samples often found in marketing research could hurt the external validity of projective techniques (Kassarjian, 1974; Donohue, 2000). In a mixed methods setting, it may be that projective techniques can be used to help elicit new ideas and hypotheses, which could be tested with empirical methods with rigorous metrics on larger samples (Donohue, 2000; Webb, 1992). In fact, there is a research culture in mixed methods research where qualitative methods are used to gain greater understanding and depth of insight into a research domain, and quantitative methods, which include rigorous measures of reliability and validity, are used to develop constructs and test hypotheses (e.g., Harrison III, 2013; Leech et al., 2010; McKim, 2017). Antunes et al. (2024) examined projective techniques from a qualitative prism and suggested utilizing qualitative evaluation of research, and adapting general work on evaluating qualitative methods (Lincoln & Guba, 1986; Shenton, 2004) They described a set a qualitative evaluation methodologies for projective techniques. These include the credibility of the research, the transferability to other contexts, the dependability of the research, and the confirmability. These metrics are inferred using methods such as research audits, detailed observation of respondents, subject and peer debriefings, and triangulation with other data sources and research.

To summarize, the evaluation of research methods is an important part of the research process. When using projective techniques in a clinical setting, the calculation of quantitative metrics for reliability and validity are an important part of establishing the correctness of the techniques. In a consumer research setting, the processes are less definite, with arguments for both quantitative reliability and validity analysis and more nuanced qualitative evaluative criteria. We explore these issues further when reviewing and evaluating academic consumer research that utilizes projective techniques.



# A REVIEW OF PROJECTIVE TECHNIQUES IN CONSUMER RESEARCH

In this section, we review the use projective techniques in academic consumer research. There are several reasons for the review. First, to understand the scope of projective techniques in consumer research. Second, to understand the different methodological traditions used when implementing projective techniques in consumer research. Third, given the mixed methods focus of this article, to understand how quantitative and qualitative methodologies are integrated when using projective techniques. The review is systematic in that a systematic process was used to find candidate articles but to keep the review manageable, some judgment was used to select the final set of articles. Using the categorization of Paré et al. (2015), the review also has critical (contrasting the different approaches to projective techniques and evaluating the use of quantitative methods) scoping (analyzing the scope of projective techniques), and descriptive (summarizing the use of different methods) elements.

Given the broad focus on consumer research in this article, a wide net was cast when looking for articles. Both the marketing and hospitality & tourism disciplines have work on projective techniques, though in both these disciplines projective techniques form quite niche subareas of general work in qualitative research. For example, in marketing, a good proportion of projective technique articles are found in the journals Qualitative Market Research and International Journal of Market Research (Campos et al., 2020), which are more open to qualitative research than many of the mainstream quantitative-focused marketing journals. There is also a strong tradition of using projective techniques in consumer research in the consumer & food science area. This research is predominantly from Latin America and tends to be more quantitatively focused than research in marketing and hospitality & tourism, with more use of statistical



analysis and a focus on word association tasks. An excellent review of word association work in this area is given in Rojas-Riva et al. (2022).

Given the breadth of the review and the detailed analysis of individual articles, a complete population sample of articles was not feasible. Thus, a stratified purposive sampling (Campbell et al., 2020) approach was used, selecting n = 25 articles from each subdiscipline. Articles were selected from a candidate pool of articles. These were selected from previous reviews (Campos et al., 2020; Rojas-Rivas et al., 2020) and via a Scopus search. Scopus is widely used for bibliographic analysis (Mongeon & Paul-Hus, 2016). Scopus searches were used for the terms "projective techniques" and "projective methods" and also for the individual types of technique (e.g., "word association" and "sentence completion". Searches were performed for general business and science categories and then articles not in any of the three subdisciplines were removed (e.g., organizational psychology projective technique articles). Articles were selected to give a good spread of projective techniques and a mix of topics and authorship groups. For example, where an author had multiple papers using a uniform methodological approach, only one of these papers was selected. A broad definition of projective techniques was used, encompassing the ordering and arrangement, association, completion, construction, and creation/expression categories described earlier. However, pure ordering and arrangement articles were not included, as this type of work has traditionally come under the banner of quantitative psychometric work and has a long history in quantitative marketing (e.g., Carroll & Green, 1997; France, & Ghose, 2019).

To be included, the papers needed to have a clear descriptions of the implementation of the projective technique, the sampling procedure, and the method of data analysis. For hospitality & tourism, the sample is almost a "population sample", as less than 30 articles were found. The



purposive sample for the consumer & food science category is much more selective, given the relative popularity of projective techniques in this field. For example, the review by Rojas-Rivas et al. (2020) found over 70 papers, just for word association tasks. Marketing falls somewhere between these poles, with 25 papers selected from a population of around 40 papers.

For each paper, the Appendix Table A1 contains the citation, a summary of the objects(s) of interest, the projective category or categories the paper falls under, a brief description of the projective stimuli, a description of the methods used to process the data, and a description of the sampling procedure. If other methods are used as part of the analysis, these are described, along with the methods used to analyze the data. Methods utilized to analyze the projective data are distinguished using bold text. As noted in Pich et al. (2015), in consumer research contexts, there is a strong overlap in definitions between the creation and construction categories. For example, one could consider the act of writing a play or advert either a creative or a construction task. Thus, creation and construction categories are merged into a single category. When projective and non-projective techniques were combined in one paper, the sample size was given for the projective techniques. Several papers include multiple implementations of projective techniques with different sample sizes. In these cases, the median sample size for the projective techniques was taken. Overall, the median sample sizes were 160 for consumer & food science, 30 for hospitality & tourism, and 50 for marketing.

Several insights can be gleaned from the analysis. First, marketing and hospitality & tourism researchers mostly use projective techniques as traditional qualitative method,s with smaller sample sizes, no use of statistical techniques, and qualitative research processes involving gaining deep understanding of specific research contexts and a focus on insights from individual subjects (e.g., Lim, 2024; Sergi & Hallin, 2011). Many of these papers utilized direct quotes



from individual responses to build arguments and conclusions. To give a few examples, Ulrich and Tissier-Desbordes (2018) utilized a small, non-probability snowball sample of n = 20 and build up arguments for "resistant masculinity" among male consumers using quotes from subjects discussing masculine and feminine brand traits, and Yam et al. (2017) utilized a small non-probability sample of n = 17 and used quotes from consumers examining how household behavior was impacted by the use of mobile game designed to encourage sustainable behavior. Second, tied to the purer qualitative focus in marketing and hospitality & tourism is the greater use of creative projective techniques. Figure 1 displays the number of association, completion, and creation implementations for each of the three source domains[1]. As some papers implemented multiple techniques, the overall count of techniques employed is greater than the number of papers.

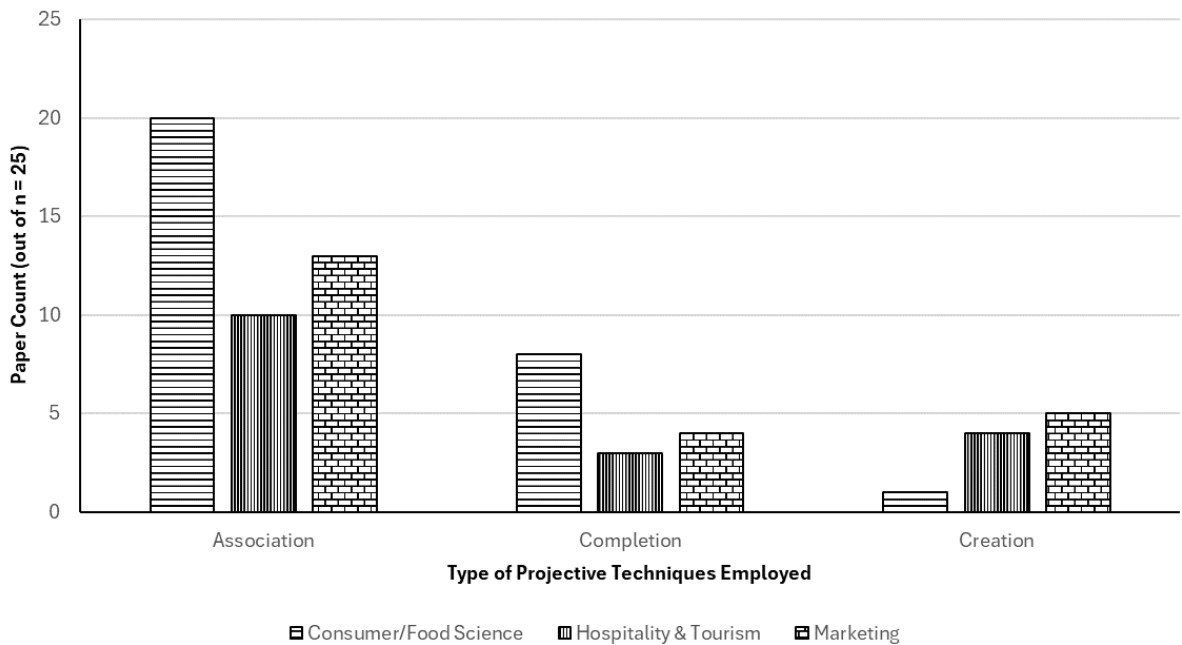

**Figure 1: Type of Projective Technique by Subdiscipline**

---

[1] Note: As the sample of papers was not a pure probability sample, we perform a descriptive analysis, but do not utilize statistical tests.



Consumer & food science has more association and completion papers than the other areas. This area has a general focus on methods with simpler data structures that can be categorized and quantified. There are numerous papers that include both association and completion tasks to elicit consumer preferences. For example, Eldesouky et al. (2015) had consumers give word associations for cheese packages and then supplemented this with completion exercises, with pictures and sentence completion bubbles related to cheese packages. The marketing and hospitality & tourism areas both have more pure creation exercises than consumer & food science. These creation exercises ranged from story creation tasks (e.g., Little & Singh, 2014; Kim et al., 2016) though to complex modeling exercises, such as an implementation of the village test (Mabille, 1950), where respondents built a model village and placed their house, an enemy's house, and different service brands in the village (Mzahi, 2014). There were a number of papers in the hospitality & tourism sample that utilized the Zaltman Metaphor Elicitation Technique (ZMET) (Zaltman & Zaltman, 2008). ZMET is a structured multi-step process to elicit deep understanding of the research domain and depending on the variant, includes six to eight steps, including picture selection, collage creation, creative description, sorting, metaphor elaboration, and construct elicitation tasks. Example domains in the selected papers include embodied hospitality experiences (Ji & King, 2018), fears and risks of pandemic travel (Jung, 2022), and hotel brand equity (Kim & Kim, 2007).

A major focus of this analysis is on how quantitative methods can be used to add additional analysis insight to projective method implementations. There are two types of implementations of quantitative methods in the sampled papers. The first is the use of quantitative methods to analyze data from projective techniques. This use of projective methods was found mostly in the consumer & food science papers, where quantitative methods were used to help categorize and



analyze data gained from projective techniques. The second use of quantitative methods is to combine a quantitative survey or other data collection method with the projective methods and analyze the quantitative data with quantitative methods. This, traditional use of mixed methods can be used to gain complementary insights, give a complete picture of the study domain, corroborate findings from the qualitative methods (and vice-versa), utilize the strengths of both types of methods, and gain a diversity of perspectives (Venkatesh et al., 2013). Examples of this use of quantitative methods include combining projective word associations with quantitative psychological distance measures (Hilverda et al., 2016), combining projective sentence completion for CRM scenarios with a quantitative Likert scale survey of views of e-CRM (Harrigan et al., 2016), and combining projective association and completion analysis of buyer and product characteristics with a quantitative willingness to pay measure (Pinto et al., 2018).

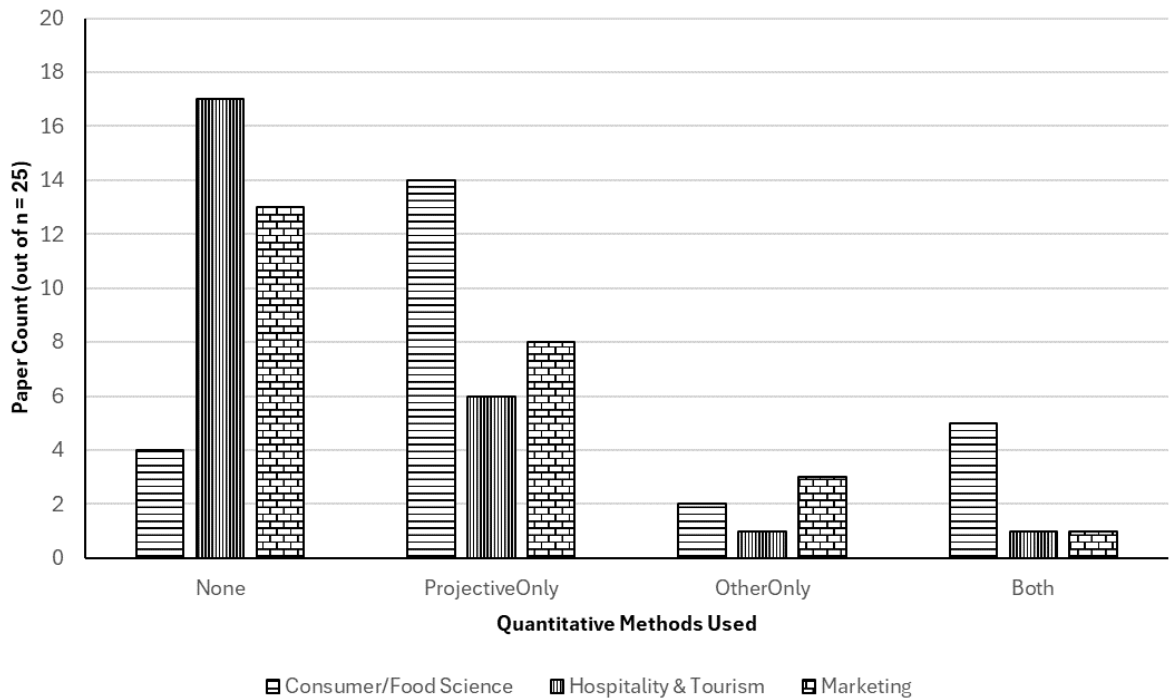

**Figure 2: Quantitative Methods Used by Subdiscipline**

Figure 2 summarizes the use of mixed methods in the sampled literature. The majority of marketing articles (n = 17) and hospitality & tourism articles (n = 13) are pure qualitative



research, while the majority of the consumer & food science articles (n = 21) include quantitative

analysis on either projective technique data, on other data, or on both.

## QUANTITATIVE METHODS REVIEW AND ANALYSIS

To further understand the use of quantitative methods, the quantitative methods used across all

sampled n = 75 papers were categorized and counted. Summary data are given in Table 1. The

method counts are split by the use of quantitative methods for projective technique data or for

other separate (mostly survey) quantitative data. The quantitative methods used on other

quantitative data encompass a standard set of techniques for analyzing survey data, including

tests for reliability and validity of scales (e.g., Cronbach's α and factor analysis), methods for

comparing means (e.g., two sample tests and ANOVA), through to more complex methods for

testing models, such as partial least squares SEM.

**Table 1: Quantitative Methods on Projective Data and Other Data**

| Quantitative Methods on Projective Data | Quantitative Methods on Other Data |
|---|---|
| Chi Squared Test of Association (18) | Cluster analysis (3) |
| Correspondence Analysis (13) | Two sample mean tests (3) |
| Chi Squared Test: Cellwise (8) | ANOVA (2) |
| Cluster Analysis (7) | Factor Analysis (2) |
| Chi-Squared Test (One-way) (3) | Multidimensional Scaling (2) |
| Coder Agreement: Cohen, Krippendorff, etc. (3) | Partial Least Squares SEM (2) |
| Cognitive Saliency Index (3) | Reliability (Cronbach's α) (2) |
| Cochran's Q (1) | Conjoint Analysis (1) |
| Correlation (1 | Correlation (1) |
| Fisher Exact Test (1 | MANOVA (1) |
| General Linear Model (likelihood of category) (1) | Principal Components Analysis (PCA) (1) |
| Latent Class Analysis (1) | |
| Multiple Factor Analysis (1) | |
| Partial Least Squares Discriminant Analysis (1) | |
| Perceptual Map (general) (1) | |
| Quantitative Measures of Association (1) | |
| Regression (1) | |
| Sentiment Analysis (1) | |



A major focus of this article is on the quantitative methods used to help analyze projective technique data. As most data generated by projective techniques are unstructured text or image data, which are not amenable to quantitative analysis of numeric data, before employing quantitative techniques, some pre-processing of data is required. This can be done through a process of content analysis to elicit categories and categorize responses or through automated data analytic techniques (e.g., Boumans & Trilling, 2018; Hagen, 2018), such as cluster analysis and LDA (latent Dirichlet analysis). There was little use of automated methods in the sampled papers. An exception is a paper that utilized computational sentiment analysis to examine the polarity of emotions in responses (Rocha et al., 2018)

In the sampled literature, nearly all the papers utilizing quantitative methods on the projective data utilized some form of content analysis to categorize data. Content analysis can be both inductive (looking bottom-up from the data to find patterns and inferences and deductive (using current theory and domain knowledge to build categorizations) (Kuckartz & Rädiker, 2023, pp. 49-76). There are no strong patterns in the use of content analysis and many papers freely mix inductive and deductive approaches, but there is more focus on inductive content analysis in consumer & food science and more on deductive content analysis in marketing. This may be because of the different focus on contributions, with more practical contributions in consumer & food science, with work on how consumers evaluate specific food products, versus more focus on building generalizable theory in marketing, for example, utilizing previous theory on consumer-brand relationship constructs to brand attachment scenarios (Cui et al., 2018). Content analysis can encapsulate a pure qualitative approach or a mixed approach involving the calculation of quantitative metrics, particularly metrics of inter-coder reliability, such as the previously described Cohen's kappa and Krippendorff's alpha. Surprisingly, only three of the



reviewed papers reported these reliability statistics and most papers did not, even when using the categorizations for quantitative analysis.

Overall, for the proactive technique data, the most commonly used method was the chi-squared test and related Fisher's exact test. Commonly used methods of structuring data included correspondence analysis and cluster analysis. Three papers (Alencar et al., 2021; Ares et al., 2014; Rojas-Rivas et al., 2020) used a cognitive salience index (CSI) (Sutrop, 2001), which can be used when multiple associations are listed by subjects, and gives a measure of the salience of a category in the mind of consumers. The index is given in (1).

$$CSI_j = \frac{n_j}{N\bar{R}_j} \tag{1}$$

, where $CSI_j$ is the cognitive salience index for category $j$, $n_j$ is the number of mentions of category $j$, and $\bar{R}_j$ is the rank for category $j$. There are also several methods, binary-response GLM (general linear model) and PLS (partial least squares) discriminant analysis, where category responses were used to predict binary indicator variables. For example, in a study of attitudes to breast feeding and baby formula (Ares et al., 2020), binary-response GLM was used to compare responses for two sub-samples, health professionals and mothers.

Further discussion and analysis of the most commonly used methods (chi-squared and related tests, correspondence analysis, and cluster analysis) are given in subsequent subsections.

**Chi-Squared Test of Association: Discussion and Reanalysis**

The chi-squared test is the most commonly used statistical test found in the literature sample. The most common implementation of a chi-squared test is the basic chi-squared test of association. Here, one is looking for evidence to reject the null hypothesis that there is independence (no association) and find that there is an association (or dependence) between two sets of categories. Here, each set of categories is a partition of the items being analyzed. In the



analyzed papers, this type of test is typically used to test for association between the categories extracted from the content analysis and some respondent demographic characteristic or product or service characteristics. The tests with respondent demographic characteristics are often used as a test of validity, checking if demographic groups that are expected to exhibit different behavior differ with respect to the categories. Examples include respondent differences with respect to language (Little & Singh, 2014), gender (Pacheco et al., 2018), country (Kim et al., 2016), region (da Silva et al., 2021), and loyalty segment derived from a quantitative survey (Doherty & Nelson, 2008). The tests of categories with respect to product characteristics include different sets of product or service attributes including mung bean foods (Dahiya et al., 2014), milk products (Esmerino et al., 2017), store type (Broeckelmann, 2010), eggs (Sass et al., 2018), fermented milk products (Pinto et al., 2018), sausage (Rocha et al., 2018), and pizza (Pontual et al., 2017).

The chi-squared test is a parametric approximation to the Fisher exact test. A p-value for the either test corresponds to the proportion of the results more extreme (further away from independence) than the result obtained. The chi-squared test statistic is given in (2). The Chi-squared distribution is a distribution of the squared deviations from the mean of $k$ independent Gaussian variables, where $k = (r - 1)(c - 1)$, $r$ is the number of rows, and $c$ is the number of columns.

$$\chi^2 = \sum_i^r \sum_j^c \frac{\left(O_{ij} - E_{ij}\right)^2}{E_{ij}} \quad E_{ij} = \frac{R_i C_j}{N} \tag{2}$$

The chi-squared statistic is given in (2), where $O_{ij}$ is the observed value in the contingency table for row $i$ and column $j$, $E_{ij}$ is the expected value, $R_i$ is the row total for row $i$, $C_j$ is the column total for column $j$, and N is the total sum of values in the contingency table. Larger values of $\chi^2$



indicate stronger association and give smaller p values. The $\chi^2$ statistic is distributed with a $\chi^2$ distribution with $k = (r - 1)(c - 1)$ degrees of freedom. Given the potential for erroneous results with NHSTs (null hypothesis significance tests) and the need for additional evidence to help validate results from statistical tests (Harvey, 2017; Schwab, et al., 2011; Wasserstein & Lazar, 2016), effect sizes (Cohen, 1988) can be used to give a sample-size independent measure of the strength of effect. Effect sizes defined for the chi-squared test include Cohen's $\omega = \sqrt{\chi^2/N}$ and Cramer's $V = \sqrt{\chi^2/(N \times min\{r - 1, c - 1\})}$. Cohen's $\omega$ is invariant to the size of the contingency table and has fixed cutoffs for small (0.1), medium (0.3), and large (0.5) effect sizes, while Cramer's V accounts for the size of the table, but the effect size cutoffs depend on the degrees of freedom.

The chi-squared test has certain assumptions, outlined in most applied statistics textbooks (e.g., Camm et al., 2024, pp. 571-577). These include a random sample of population, independence of observations and mutually exclusive categories, and a minimum value of $E_{ij}$. This "minimum" is usually set at 5, but authors have critiqued this criterion and have offered alternatives, particularly for $2 \times 2$ contingency tables (e.g., Andrés & Tejedor, 2000).

To further understand the application of Chi-squared (and other tests) we gathered all contingency tables that were contained in the literature sample. Most were utilized for either chi-squared tests of correspondence analysis (or both), though a few were utilized for exploratory analysis. Some papers included multiple contingency tables and some tables were split by a third categorical factor, for example, clusters for healthy eating preferences (Viana et al., 2014). In most cases, at least one of the category variables was a coded category from the projective analysis. The data are summarized in Table 2. Columns are included for the paper, the row and



column category variables, along with the number of categories for each variable, and any

categorical splits, along with the number of categories for each split variable.

**Table 2: Datasets for Reanalysis of Quantitative Data**

| Paper | Row | nRow | Column | nCol | Split | SplitName |
|---|---|---|---|---|---|---|
| Broeckelmann (2010) | Coded Cat. | 6 | StoreType | 4 | | |
| Boreckelmann (2010) | Coded Cat. | 6 | Price | 3 | | |
| da Silva et al. (2021) | Coded Cat. | 10 | Region | 5 | | |
| da Silva et al. (2021) | Coded Cat. | 8 | Region | 5 | | |
| Dahiya et al. (2014) | Coded Cat. | 11 | Food | 15 | | |
| Doherty & Nelson (2008) | Coded Cat. | 2 | LoyaltySegment | 3 | | |
| Doherty & Nelson (2008) | Coded Cat. | 7 | LoyaltySegment | 3 | | |
| Esmerino et al. (2017) | Coded Cat. | 15 | MilkProduct | 8 | | |
| Gambara et al. (2014) | Coded Cat. | 14 | Cosmetic Type | 5 | | |
| Hilverda et al. (2016) | Coded Cat. | 29 | FoodStimulus | 3 | | |
| Hilverda et al. (2016) | Coded Cat. | 29 | OrganicUsage | 3 | | |
| Kachersky & Lerman (2013) | Sentiment | 3 | Perspective | 3 | | |
| Kim et al (2007) | Coded Cat. | 7 | Country | 3 | | |
| Little & Singh (2014) | Coded Cat. | 2 | Language | 4 | 2 | UrbanRural |
| Little & Singh (2014) | Coded Cat. | 2 | Language | 4 | 2 | Gender |
| MeliouMaroudas (2010) | Coded Cat. | 10 | Demographic | 4 | | |
| Pacheco et al. (2018) | Coded Cat. | 12 | Gender | 2 | | |
| Pinto et al. (2018) | Coded Cat. | 10 | FermentedMilk | 6 | | |
| Pinto et al. (2018) | Coded Cat. | 10 | FermentedMilk | 6 | | |
| Pontual et al. (2017) | Coded Cat. | 7 | Pizza | 4 | 2 | CeliacGroup |
| Rocha et al. (2018) | Coded Cat. | 24 | Sausage | 6 | | |
| Rojas-Rivas et al. (2020) | Coded Cat. | 18 | Age | 4 | | |
| Rojas-Rivas et al. (2020) | Coded Cat. | 18 | Education Level | 3 | | |
| Sass et al. (2018) | Coded Cat. | 11 | Egg | 6 | | |
| Sass et al. (2018) | Coded Cat. | 11 | Egg | 6 | | |
| Soares et al. (2017) | Coded Cat. | 17 | Region | 2 | | |
| Viana et al. (2014) | Coded Cat. | 19 | Category | 4 | | |
| Viana et al. (2014) | Coded Cat. | 19 | Category | 4 | 3 | HealthCluster |



| Vidal et al. (2013) | | Coded Cat. | 9 | List Type | 2 | | |

In total, with category splits included, there were n = 34 contingency table. For each contingency table, we ran the chi-square test, along with the Fisher's exact test, both using the base statistics package in R 4.4.1. The Fisher's exact test function utilized the FEXACT algorithm described in Clarkson et al. (1993). Given the infeasibility of the Fisher's exact test on larger contingency tables, a probabilistic non-exact variant of the test was run. In addition, for each contingency table, statistical power was calculated for small ($\omega = 0.1$), medium ($\omega = 0.3$), and large ($\omega = 0.5$) effect sizes, using the *pwr* package (V1.3.0) in R (Champely et al., 2020). The analysis of power is important, as when running hypothesis tests, researchers will often infer the null hypothesis when failing to reject the alternative hypothesis, but when statistical power (i.e., the probability of rejecting the null hypothesis given a set effect size or deviation from the null hypothesis) is low, these conclusions are liable to be erroneous. Many studies, even in top ranked journals, are under powered and do not include power calculations (e.g., Abraham & Russell, 2008; Cashen & Geiger, 2004; Meyners et al., 2020).

The results of the reanalysis are given in Table 3. Summary statistics are given for chi-squared value ($\chi^2$), the p value for the chi-squared test, the p value for the Fisher exact test, the proportion of cells with $E_{ij} >= 5$, the Cohen's Omega ($\omega$) and Cramer's V effect sizes, and the three power values.

**Table 3: Summary of Reanalysis of Chi-Squared Association Tests**

| Statistic | Min | 10 %ile | 25 %ile | Median | Mean | 75 %ile | 90%ile | Max |
|---|---|---|---|---|---|---|---|---|
| Chi-sq | 0.2829 | 7.7129 | 36.0541 | 95.8036 | 318.0301 | 361.2837 | 944.8105 | 2496.4226 |
| Chisq p | 0.00000 | 0.00000 | 0.00000 | 0.00000 | 0.0689 | 0.0022 | 0.2026 | 1.0000 |
| Fisher p | 0.00050 | 0.00050 | 0.00050 | 0.00050 | 0.0711 | 0.0036 | 0.2284 | 1.0000 |
| PropE≥5 | 0.0278 | 0.1431 | 0.5000 | 0.6839 | 0.6517 | 0.9226 | 1.0000 | 1.0000 |
| Omega | 0.0349 | 0.2081 | 0.4073 | 0.5542 | 0.5917 | 0.8130 | 1.0880 | 1.2183 |



| CramersV | 0.0247 | 0.1471 | 0.2414 | 0.3010 | 0.3658 | 0.5274 | 0.6085 | 0.8255 |
| PwrSmall | 0.0756 | 0.0850 | 0.1759 | 0.4245 | 0.4752 | 0.7880 | 0.9054 | 0.9995 |
| PwrMed | 0.2906 | 0.3757 | 0.9334 | 1.0000 | 0.8975 | 1.0000 | 1.0000 | 1.0000 |
| PwrLarge | 0.6500 | 0.7805 | 0.9999 | 1.0000 | 0.9683 | 1.0000 | 1.0000 | 1.0000 |

Several conclusions can be noted from the analysis. Overall, the majority of contingency tables showed significant results. The Fisher's exact test was a little more conservative, but not hugely more conservative than the chi-squared test, which agrees with the past literature (Andrés & Tejedor, 1995). The 25[th] percentile value of Omega ($\omega$) is 0.4073, which is a moderate effect size, giving further evidence that most of the projective technique categorizations have some association with the product, service, or demographic variables, typically used in the second category. Overall, the power was quite high for medium and large effects, but the average power for Omega ($\omega$) = 0.1 small effects was less than 0.5, indicating that the tests would find the correct result for small effects less than 50% of the time. Overall, one should be cautious about making conclusions from non-significant test results. The mean and median proportion of cells meeting the $E_{ij} >= 5$, criterion are 0.6517 and 0.6839 respectively. The 75[th] percentile of contingency tables still has some cells not meeting the criteria. This result is probably due to i) large contingency tables relative to the number of subjects, and ii) an uneven distribution of category values.

**Cellwise Chi-squared Tests**

The chi-squared test only finds an overall association between categorical variables. To find more detailed insight, many of the papers in literature review utilized cellwise chi-squared tests. Here, for each cell in row $i$ and column $j$ in the contingency table, a new $2 \times 2$ contingency is created, by combining the remining $i - 1$ rows into a second row and $j - 1$ columns into a second column. Each individual chi-squared test is carried out on a new $2 \times 2$ contingency table, in most



cases using the Yates continuity correction, which is recommended for $2 \times 2$ contingency tables (e.g., Yates, 1984). In most cases, significant results are marked with a (+) or a (-), depending on whether the observed value is more or less than expected. To understand these tests a little better, we repeated the process used for the chi-squared test on the full contingency table with cellwise chi-squared tests. The only differences were the use of the continuity correction and the chi-squared test and the fact that both ω and Cramer's v reduce to the phi $\phi = \sqrt{\chi^2/N}$ effect size. Each contingency table resulted in $c \times r$ tests, and the mean values for each table were calculated and are summarized in Table 4. In addition, the proportions of significant results for both tests are summarized in the second and third rows. There has been much research and argument regarding the required expected value of cells for $2 \times 2$ tests. We include two additional metrics. The first, from Haber (1980), is more conservative, and sets $E_{ij} \geq max\{5, N/10\}$. The second, developed by Andrés et al. (2005) for Chi-squared tests with continuity corrections, sets $E_{ij} > 3.9$ for $N \leq 500$ and $E_{ij} > 6.2$ for $N > 500$ for the Yates continuity correction.

**Table 4: Summary of Reanalysis of Cellwise Chi-Squared Association Tests**

| Measure | Min | 10 %ile | 25 %ile | Median | Mean | 75 %ile | 90%ile | Max |
|---|---|---|---|---|---|---|---|---|
| Chisq | 0.037 | 0.950 | 1.580 | 3.692 | 7.585 | 8.400 | 17.665 | 71.824 |
| PSig | 0 | 0.0238 | 0.0833 | 0.2822 | 0.3080 | 0.5401 | 0.6667 | 0.6970 |
| PSigFisher | 0 | 0.0238 | 0.104 | 0.3046 | 0.3293 | 0.5600 | 0.6667 | 0.7273 |
| pVal | 0.0693 | 0.1267 | 0.2049 | 0.3582 | 0.3827 | 0.5434 | 0.6595 | 0.8819 |
| Fisherp | 0.0526 | 0.1086 | 0.1777 | 0.3210 | 0.3517 | 0.5027 | 0.6441 | 0.8485 |
| Phi | 0.0088 | 0.0317 | 0.0537 | 0.0802 | 0.1234 | 0.1416 | 0.3585 | 0.5381 |
| Prop. E>5 | 0.5592 | 0.6557 | 0.7771 | 0.9005 | 0.8712 | 0.9788 | 1.0000 | 1.0000 |
| Haber | 0.3500 | 0.4594 | 0.5000 | 0.5556 | 0.5779 | 0.6272 | 0.7500 | 0.8333 |
| Andrés | 0.5987 | 0.6689 | 0.8333 | 0.8857 | 0.8759 | 0.9586 | 1.0000 | 1.0000 |
| PwrSmall | 0.0756 | 0.0850 | 0.2392 | 0.6020 | 0.5642 | 0.8908 | 0.9835 | 1.0000 |
| PwrMed | 0.2906 | 0.3757 | 0.9629 | 1.0000 | 0.9103 | 1.0000 | 1.0000 | 1.0000 |
| PwrLarge | 0.6500 | 0.7805 | 1.0000 | 1.0000 | 0.9686 | 1.0000 | 1.0000 | 1.0000 |



It is noticeable that only a small proportion of the tests are significant, with around 30% of tests significant. The Fisher exact test is a little less conservative than the chi-squared test. The effect sizes for the significant tests are much lower than for the overall table tests. In fact, the mean effect size of $\phi = 01234$ is small and it is not until the 90th percentile that effect sizes are medium. This aligns with the small proportion of significant tests. Given the reduced number of cells and unchanged sample size for the cellwise Chi-squared tests, a larger mean proportion of tests met the minimum sample size criteria, particularly using the standard $E_{ij} \geq 5$ criterion (0.871) and the Andrés et al. (2005) criterion (0.876). This value was lower (0.578) for the more conservative Huber (1980) criterion.

**Chi-squared Test Conclusions**

The chi-squared test has been utilized as the primary method of statistically analyzing data derived from projective techniques. The use of chi-squared tests for association analysis has been used to help validate categorizations of qualitative data collected using projective techniques and used to examine the relationships between projective categories and a range of product, service, and respondent characteristics. However, there needs to be several notes of caution as to the implementation of these tests.

**General Testing:** Chi-squared tests and associated tests assume a random sample from the population. In most cases found in the literature review, the sampling procedure is a typical qualitative purposive or convenience non-probability sample. In such cases, it is difficult to extrapolate results of the research to the entire population, particularly in quasi-experimental designs, which do not account for potential control and unobserved variables. While one could assume that the categorized associations and completions are generated from a random process



and these categorizations are relatively consistent across sub-populations, one still needs caution when interpreting chi-squared statistics.

**Independence Assumption:** In many of the contingency tables (about 50% of the sample), the number of observations is greater than the number of respondents. Again, one could consider a "response" from an association task to be the tested random process, but when a single respondent generates multiple categories, this in some sense violates the independence and mutual exclusivity assumptions of the chi-squared test.

**Sample Size Assumption:** Many of the contingency tables have cells with expected values less than 5 (or the alternative criteria for the $2 \times 2$ tests). This in part is due to the large contingency tables generated by categorizations of projective technique data. Though the chi-squared test may be robust against slight deviations from this requirement (Koehler & Larntz, 1980), given that the Fisher exact test has computationally efficient algorithms for dealing with large contingency tables (e.g., FEXACT), and as noted in the reanalysis, is only slightly more than the Chi-squared test, in situations where expected values are small, it would be a good idea to run the Fisher exact test, either as a replacement for or as a supplement to the Chi squared test.

**Effect Size and Power:** There is virtually no use of effect size and power calculations in the reviewed literature. When test assumptions are violated and it is difficult to extrapolate sample findings to a population, effect sizes can give a good estimate of the overall magnitude of effects found in the specific samples. In some cases, reviewed papers made conclusions about the lack of effect without considering statistical power. Such conclusions may be erroneous without power calculations to determine the likely power and Type II error.

**Content Analysis and Categorization:** While using a purely qualitative process and qualitative content analysis of categories is a valid research choice, if utilizing categories in statistical



analysis, some measure of coding reliability, such as Cohen's kappa or Krippendorff's alpha, could help understand and quantify measurement error. This is important, as measurement error can affect statistical estimates, reducing the efficiency of parameter estimation (Bound et al, 2001) and potentially introducing bias (Hyslop & Imbens, 2001).

**Cellwise Chi-squared Test:** The use of cellwise chi-squared calculations was common in the sampled paper. There is strong logic behind the use of such techniques, as the chi-squared test is an "omnibus" test with respect to the categories and does not give inference as to the relative level of association between individual categories. However, strong caution must be applied when running cellwise calculations, particularly when designating each cell as significant or non-significant. The first four points hold and in addition, without multiple comparisons correction (e.g., Tukey, 1991), analyses will suffer from compounded Type I error (or Type II error for non-significant results). There are numerous methods of calculating and visualizing cell scores (e.g., Mirkin, 2024). One metric is to calculate the percentage increase in observed value over expected, $100 \times \left( O_{ij} - E_{ij} / E_{ij} \right)$, which is related to the chi-squared distance criterion used in correspondence analysis. An example of this metric is given for the Pinto et al. (2018) paper for a Haire's shopping list exercise for different fermented milk products. One can see a clear association between discussing health disorders and lactose free products and discussing being on a diet and light products.



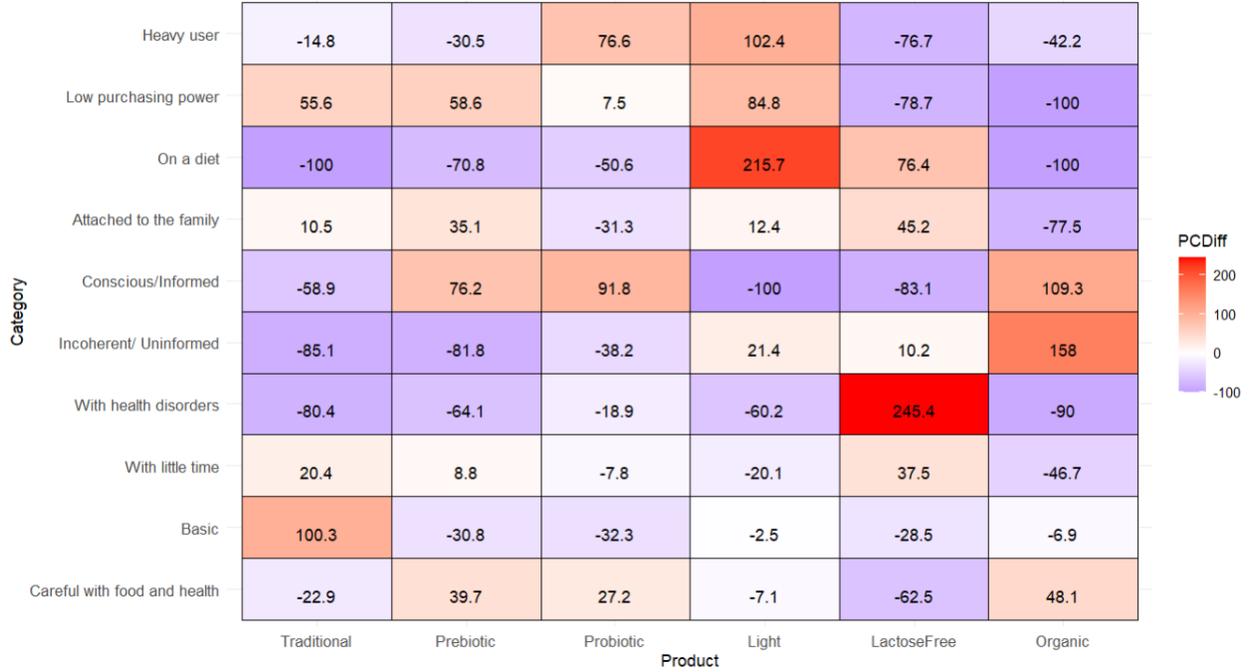

**Figure 3: Pinto et al. (2018) Shopping List %age Difference from Expected Heatmap**

**Data Structuring Approaches: Correspondence Analysis and Cluster Analysis**

After chi-squared tests, the most common methods of analyzing the projective technique data were correspondence analysis and cluster analysis. Each of these methods are used to help explore the structure of the data and relationships between categories and products.

Correspondence analysis is a method that allows both row and column categories to be analyzed and plotted on the same map. The full detail and mechanics of correspondence analysis are given in Greenacre (2017).

$$s_{ij} = \frac{O_{ij} - E_{ij}}{\sqrt{NE_{ij}}} \qquad (3)$$

Correspondence analysis utilizes a chi-squared distance. First a matrix $\mathbf{S}$ is calculated, with entries $s_{ij} \in \mathbf{S}$ defined as the deviations from probabilistic independence standardized using the square root of the row and column frequencies. Using the notation introduced for the chi-squared test, this is given in (3). A singular value decomposition is then performed in a similar manner to



principal components analysis (e.g., Shlens, 2014), so that $\mathbf{S} = \mathbf{U\Delta V}^T$, where $\mathbf{U}$ and $\mathbf{V}$ are the matrices of left and right singular vectors and $\mathbf{\Delta}$ is a diagonal matrix of singular values. The row category coordinates are defined in (4) and column category coordinates are defined in (5).

$$\mathbf{X_R} = \mathbf{D_R^{-1/2} U\Delta} \qquad (4)$$

$$\mathbf{X_c} = \mathbf{D_c^{-1/2} \Delta V} \qquad (5)$$

, where $\mathbf{D_R}$ is a diagonal matrix of the marginal row probabilities, i.e., the i[th] diagonal entry is $R_i/N$, and $\mathbf{D_C}$ is a diagonal matrix of the marginal column probabilities. The variance explained by dimension $i$ is the scaled squared singular value for that dimension, i.e., $(\Delta_{ii})^2 / \sum_{j=1}^k (\Delta_{ii})^2$ for a $k$ component solution. The row and column coordinates can be combined onto a single biplot. The distances between the biplots can be interpreted as the chi-squared distances between categories, allowing a visual representation of associations. These associations can be used to analyze relationships between individual categories in a similar manner to the cellwise Chi-squared tests. Consider two examples from the reviewed papers. The Pinto et al. (2018) paper was previously used in Figure 3 to analyze the strength of individual category associations. Correspondence analysis biplots for a Haire shopping list task and a word association task are given in Figure 4. Here, the correspondence analysis can show similar associations to Figure 3. For example, in the shopping list biplot, traditional products are related to low purchasing power and lactose free products are related to health disorders. Another use of correspondence analysis is to compare solutions from different techniques as a test of consistency. For example, in the word association task, prebiotic and probiotic products are close to one another on each plot.



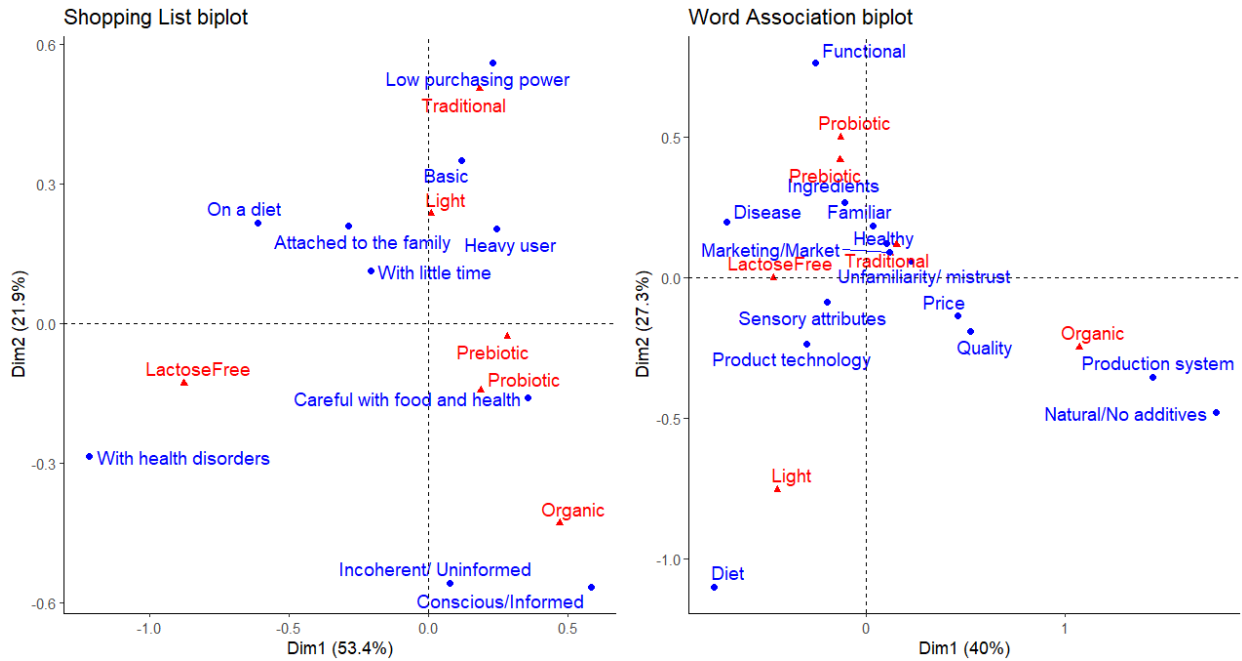

**Figure 4: Biplots for Pinto et al. (2018)**

A similar analysis is performed for Sass et al. (2018) on a word association task for different types of egg. Respondents were asked for negative and positive associations and these were analyzed separately. The resulting correspondence analysis biplots are given in Figure 5. The positive association plot accounts for 85.5% of the variance in the data and the negative association plot accounts for 75.8% of the variance in the data. Here one can see strong differences between the associations, as would be expected by the domain. For example, factory farmed white eggs have negative associations with the creation system and quality, while they have positive associations with price.



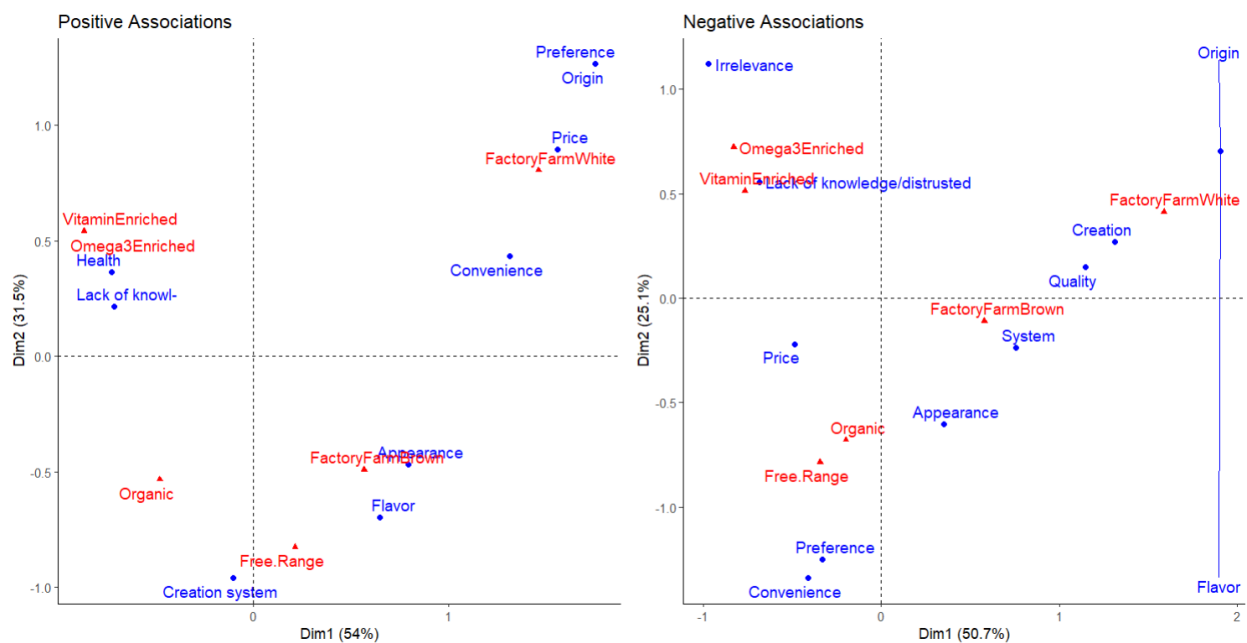

**Figure 5: Biplots for Sass et al. (2018)**

In both of the above examples, the relative positions of the projective categories and the substantive projective categories were used to help analyze similarities and differences between i) different data collection methods and ii) positive/negative associations. While in both cases, the category sets are different, the relative positions of the substantive categories can be compared. Procrustes[2] analysis can be used to test the consistency of different data configurations (e.g., Dijksterhuis & Gower, 1991; El Ghaziri & Qannari, 2015). In Procrustes analysis of two configurations (Gower, 2010), one configuration is transformed onto the second matrix using an orthogonal rotation matrix $\mathbf{R}$ (i.e., $\mathbf{R^T R = 1}$), with R set to minimize the differences between the transformed matrix and the second configuration. The VEGAN package in R (Dixon, 2003) contains the *procrustes* and *proctest* functions for implementing Procrustes analysis. The *procrustes* function contains a post-transformation measure of the sum of squared differences between configurations standardized between 0 and 1. The *proctest* function contains

a nonparametric statistical test comparing the sum of squared differences from the rotated solution with random matrix rotations. A p value < α indicates evidence of similarity between configurations, i.e., the p-value is the proportion of random matrices more similar to the source configuration under rotation. For the Pinto et al. (2018) solutions, the standardized sum of squares differences is 0.7699, with a p value of 0.5806, indicating that though there is some similarity, the two analyses generate different insights. For the Sass et al. (2018) solutions, the standardized sum of squares differences is 0.0958, with a p value of 0.0056, indicating that though positive and negative associations generate different categories, they show consistency with respect to the underlying brands.

An alternative to Procrustes analysis, is to compare the agreement between item neighborhoods. For example, in Figure 4, the Probiotic and Prebiotic products are the nearest neighbors of one another on both of the maps, so show agreement. This agreement can be aggregated over items and neighborhood sizes. This type of agreement rate has been widely used to evaluate dimensionality reduction solutions (e.g., Akkucuk & Carroll, 2006; Chen & Buja, 2009). Both the agreement rate (AR) and an adjusted AR that subtracts expected random agreement are implemented in *GenAgree* in R (France & Akkucuk, 2021). These metrics are analogous to the Rand index and adjusted Rand index (Hubert & Arabie, 1985) used for cluster analysis. For the basic AR, 0 indicates no agreement and 1 indicates perfect agreement. The adjusted AR can be less than 0 when agreement is less than random agreement. For Pinto et al. (2018), the AR values are 0.6361 (standard) and 0.0903 (adjusted). For Sass et al. (2018), the AR values are 0.8500 (standard) and 0.6250 (adjusted). These results show high concordance with the Procrustes analysis results, but have easy interpretations, similar to correlations.

**Summary and Other Methods**



This section has covered both the chi-squared and correspondence analysis methods of analyzing contingency table. These methods are linked mathematically through the chi-squared distance. The chi-squared tests provide a useful measure of association between projective categories and other categorical variables, both for validating the research process and examining substantive research issues. However, certain assumptions apply to chi-squared and related tests. One contribution of this paper is to describe these assumptions and to make recommendations for implementing association tests. Correspondence analysis provides a conceptually simple method of structuring data that does not require parametric statistical assumptions. Another contribution of this paper is to summarize and elocute ways of using correspondence analysis to gain insight from projective technique data and to compare different correspondence analysis configurations.

There are a wide range of other data structuring approaches that can be used to analyze contingency papers. Some of these are summarized in the review, but others have not yet been implemented by consumer researchers. For example, cluster analysis can be utilized to look at relationships between categories and between category configurations. There are a wide range of cluster analysis algorithms (e.g., Jain, 2010; Kaufman & Rousseeuw, 2009; Kettenring, 2006). Given space limitations and the breadth of the topic, this topic cannot be explored in depth. However, cluster analysis can be utilized in a similar way to correspondence analysis by examining relationships between categories and between different analyses. For example, Figure 6 compares hierarchical cluster analysis for the positive and negative associations gathered from Sass et al. (2018). As per the Procustes analysis and the agreement metric, there is a strong similarity in the representation of products, though for the negative associations the factory farm white eggs were grouped differently from the positive associations.



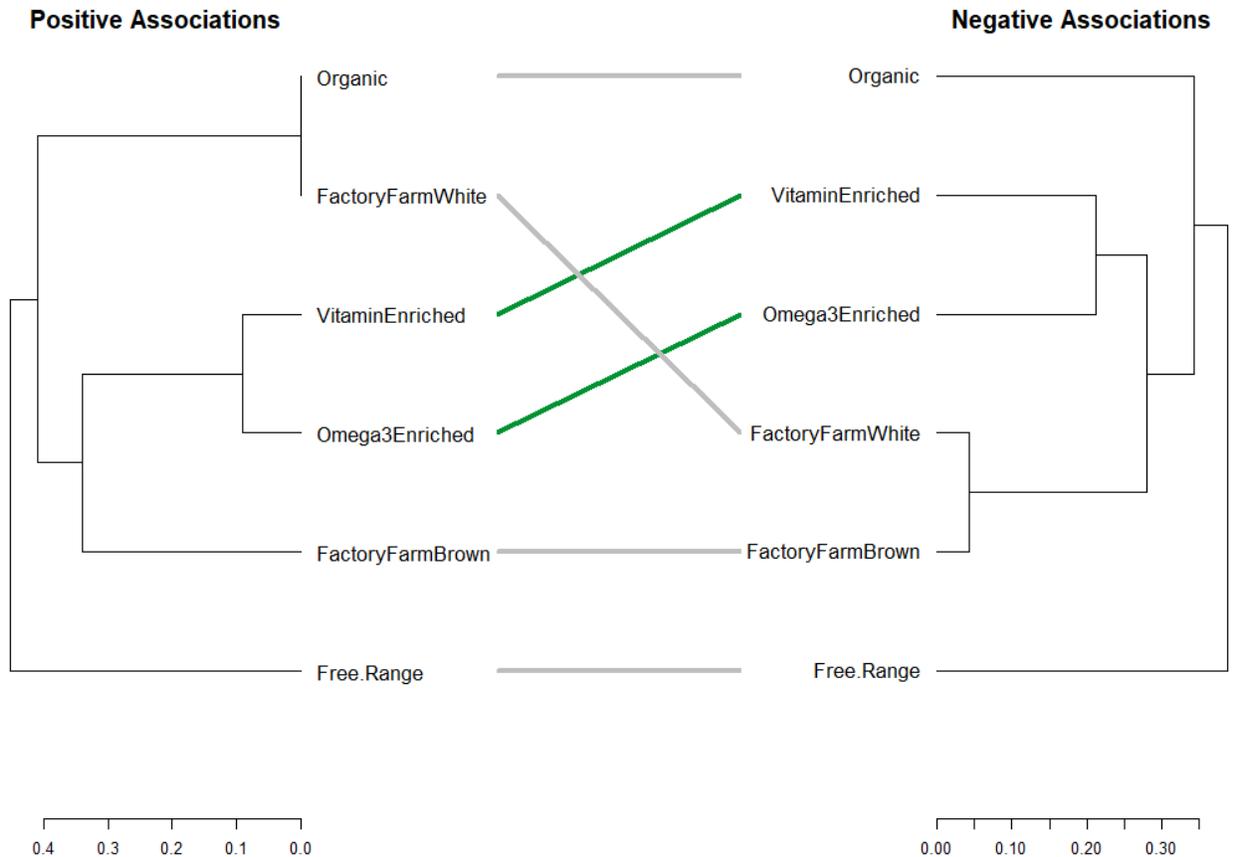

**Figure 6: Sass et al. (2018) Tanglegram of Positive Associations vs. Negative Associations**

A range of other methods could be employed to analyze the contingency table data. For example log-linear models have been widely used to model category counts in contingency tables (Agresti, 2012), are part of the general family of GLM (generalized linear models), and can be used to model complex situations such as multi-way tables and contingency tables that change over time. Class prediction methods, such as discriminant analysis can be used to help predict product and subject characteristics, e.g., gender (Pacheo et al., 2018). Overall there is scope for further improving analysis using existing methods used with projective technique data and scope for implementing new methods.

## CONTRIBUTIONS TO MIXED METHODS RESEARCH



Projective techniques span a range of scenarios ranging from single ordering of stimuli to creative tasks. Much mixed methods research has quite defined boundaries, with data gathered separately and insights from qualitative analysis and quantitative analysis triangulated and combined in the results (e.g., Guest, 2013). However, in a fully mixed design, qualitative and quantitative research components can be mixed in at different stages of the research design (Leech & Onwuegbuzie, 2009).

What makes projective techniques particularly interesting from a mixed-methods perspective is that it is difficult to define "projective research methods" as either qualitative or quantitative. In effect, a projective technique can be a "chimera", with some groups of researchers treating the methods as qualitative and others as quantitative. While some researchers treat projective techniques as qualitative and integrate with quantitative survey research in a typical mixed methods setting, others introduce utilize projective techniques as a quasi-quantitative method. However, there are dangers in that approach. Mixed methods research often needs to deal with sampling complexities (Morse, 1991; Teddlie & Yu, 2007), for example, generating a purposive qualitative sample from a quantitative sample. However, many of the methods outlined in the literature review utilized qualitative style samples and then make statistical population inferences from the data. In addition, statistical tests rely on certain assumptions, such as independence, which may not be present.

With respect to mixed methods research, this paper has several major contributions. First, the paper gives a broad-scope overview of the use of projective techniques. A literature review with stratified purposive sampling is used to compare and contrast the use of qualitative and quantitative methods across the subdisciplines of marketing, hospitality & tourism, and consumer & food science. Second, the paper is designed to be a resource for researchers



implementing projective techniques in the broad area of consumer research. While being respectful of the advances made by researchers implementing projective techniques, advice and ideas are given for improving analyses, for example, by discussing the use of effect sizes and power analyses in chi-squared tests and giving an easily interpretable metric for cellwise association in chi-squared tests. Third, the paper provides a base for future research. The synthesis of machine learning and text analytic techniques and traditional content analysis (e.g., Chen et al., 2018) is an active research area. The broad area of projective techniques has great potential for synthetic approaches at the boundary of qualitative and quantitative methods.

## LIMITATIONS AND FUTURE WORK

This work surveyed a wide variety of papers, in three different sub-disciplines and ran an empirical reanalysis on data extracted from the papers. However, these data are in summary contingency table format and thus have already had initial coding and summary analysis performed on the data. This is by necessity, as raw data are rarely divulged by researchers. However, there is scope to run primary data collection exercises and then examine both traditional content-based coding and machine learning-based analysis of text. As noted in the introduction, consumer researchers are increasingly utilizing machine learning technologies and hybrid semi-automated content analysis approaches. An interesting experiment would be to collect a range of data going from ordering data, through simple word associations, through to more complex sentence and paragraph data, then and look at the processes needed to help preprocess the data before running substantive analysis. This sort of analysis could also be used to further develop reliability and validity methods and metrics for projective techniques, a need noted earlier in the paper. Figure 7 gives an overview of the process of combining



algorithmically coded quantified data with manually coded data as a preprocessing step to some of the methods utilized to analyze data from projective techniques.

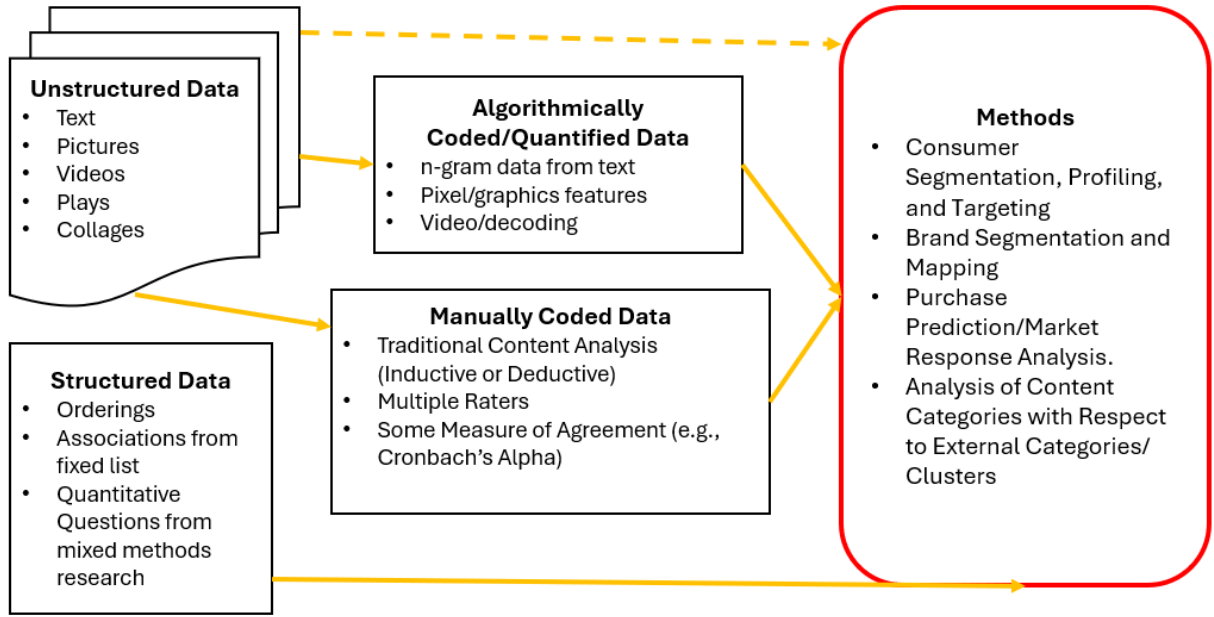

**Figure 7: Future Semi-Automated Content Analysis**

This paper focused on a range of methodologies for analyzing coded projective technique data. These included chi-squared tests, correspondence analysis, and cluster analysis. However, the techniques discussed are those found in the literature review and already widely used by researchers. There is potential to further expand the scope of techniques used in analyzing projective techniques. The methodological work in this paper could be expanded to cover additional techniques. For example, as previously noted, log-linear methods provide a consistent and flexible methodology for analyzing contingency table data. A wide range of Bayesian analysis techniques could be used to combine prior information or information from other studies into the analysis of projective techniques.

## CONCLUDING DISCUSSION



Projective techniques have a great many uses in eliciting behavioral insights and meaning from consumers. They form a vibrant, if niche set of research topics and methods. Projective techniques are somewhat at the boundary of qualitative and quantitative research and there are researchers, particularly in marketing and hospitality & tourism, who treat projective methods as "pure" qualitative research techniques. There are other researchers, particularly in consumer & food science, who treat projective techniques as "quasi-quantitative" techniques and focus on extracting structured data and quantitative insights from projective techniques. In both cases, the use of projective techniques can be combined with traditional quantitative survey research in a mixed methods context. Researchers in consumer & food science have developed innovative methods for structuring and quantitative analysis of projective data. There is huge potential for further development of these methods, both by improving rigor, improving methodology for testing reliability and validity, and incorporating modern data analytic methodologies for structuring unstructured data. With this future development, projective methods should take their place as core methods to be considered by both researchers and practitioners implementing consumer research.

**Table A1: Generative AI Performance Evaluation for Coding**

| Citation | Description | Analysis | Sample |
|---|---|---|---|
| Abalos et al. (2024) **Type**: Association, Completion **Cat:** Consumer Sci. | Word association with lasagna cooked by "sous vide" and three sentence completion tasks with picture scenarios and completion bubbles. | Coding and categorization of responses. Correspondence analysis of derived categories against the three completion scenarios. | Convenience sample of n = 198 recruited using social media. |
| Alencar et al. (2021) **Type:** Association, Completion **Cat:** Consumer Sci. | Free word association with Gluten-free (GF) bread and ordered word listings of quality attributes. Sentence completion task for scenarios involving GF bread. Combined with quantitative survey using Likert scales of the importance of bread attributes. | Content analysis to categorize terms. Correspondence analysis of categories vs. respondent consumption characteristics. Free listing terms put into a word-respondent matrix and a cognitive saliency index was calculated for the different categories. | Convenience sample of n = 205 Brazilian consumers with celiac disease. |
| Amatulli et al. (2023) **Type:** Association, Creation **Cat:** Marketing | Associate words and images related to consumption, before, during, and after the pandemic. Similarly give three consumption categories for each period. Collage construction and the creation of fairy tales related to aspects of consumption. | Qualitative analysis of responses. | Sample of n = 31 Italian undergraduate students. |
| Anghelcev et al. (2015) **Type:** Creation **Cat:** Marketing | Collect 10-12 images covering "thoughts and feelings" regarding climate change. Follow six step ZMET process, resulting in a summary collage for each subject. | Extract deep and conceptual metaphors, along with thematic categories. Qualitative discussion and quoting individuals based on metaphors and categories. | Sample of n = 12 (determined by saturation) of South Koreans, aged from 20-28 |
| Ares and Deliza (2010) **Type:** Association **Cat:** Consumer Sci. | Word association with a stimuli of label/packaging a milk drink. A free listing/association of listing expected packaging features. | Cluster analysis of categories and of subject for each experiment. Chi-squared test between category mentions and clusters. | Convenience sample of n = 200 Uruguayan consumers, with n = 100 used in each experiment. |
| Ares et al. (2014) **Type:** Association **Cat:** Consumer Sci. | Word association and free listing for consumers concepts of general and food-related wellbeing. | Content analysis and categorization. Calculation of cognitive saliency index was calculated for each of the categories. | Stratified sample of n = 120 people balanced by demographic characteristics. |
| Ares et al. (2020)<br>**Type:** Association<br>**Cat:** Consumer Sci. | Word association with breastfeeding and infant formulas. | Responses were categorized into larger categories (dimensions) and sub-categories. Responses were compared for health professionals vs. mothers across categories for breastfeeding and formula using glm. Analysis of rank vs. frequency for sub-categories. | Sample of n = 154 health professionals and cluster sample of n = 330 mothers from different health centers, giving sample overall n = 484. |
| Broeckelmann (2010)<br>**Type:** Completion<br>**Cat:** Marketing | For different retailing channel scenarios, a series of images and dialog bubbles, including special/coupon scenarios for two store types and differing prices for online purchase (less, same, or more than store price). | Use manual coding with Kendall's W consistency check to categorize the respondents' purchase intentions. Analysis of categories using the chi-squared test. | Samples from students at German university. Sizes of n = 221 (study 1), n = 150 (study 2), n = 122 (study 3), and n = 176 (study 4) |
| Celuch and Neuhofer (2024)<br>**Type:** Association, Creation<br>**Cat:** Hospitality | Open-ended questions on event behavior (at start and end of event). Space for creative drawing with sentence bubble and association of event in multiple categories (e.g., color, animal, magical creature). | Qualitative analysis, with example creative drawings and summaries of word association tasks. | Convenience sample of event attendees with n = 21. |
| Chen et al. (2022)<br>**Type:** Association, Completion<br>**Cat:** Hospitality | Word association with "service robots" and sentence completion on preferences for service robots and preferred service robot features. Combine with semi-structured interviews. | Use of coding to find themes. Qualitative analysis including quotes from individual research subjects. | Purposive sampling of generation z in China who have used service robots, with n = 33 for projective techniques and n = 15 for interviews. |
| Cherrier (2012)<br>**Type:** Construction<br>**Cat:** Marketing | Student subjects created collages of healthy and unhealthy identities, along with text descriptions. They then created colleges of student life, with text descriptions. | Qualitative analysis of text. Classification of collage into healthy and unhealthy identities. | Student sample from marketing classes of n = 53 students. |
| Chhabra (2012)<br>**Type:** Completion<br>**Cat:** Hospitality | Sentence completion tasks asking respondents to outline leisure and holiday plans in different scenarios. | Qualitative analysis of responses. | Convenience sample of n = 200 students from an American university. |
| Cian & Cervai (2011)<br>**Type**: Association<br>**Cat:** Marketing | Select brand associations photos from a photoset consisting of a range of categories (animals, houses, etc.) and then write adjectives associated with the | MuSeS qualitative analysis. Use the similarities between adjectives to reduce adjective set and use Flexigrid cluster analysis to cluster adjectives into | A sample of n = 30 randomly invited from consumers of a sailboat factory. |
| | selected pictures. Repeat process with colors, touch (of materials), and hearing. Ask the respondent for similarities between adjectives. | categories. Compare with personality traits from Aaker's scale. | |
| Cui et al (2018) **Type:** Completion, Association, Ordering, Creative **Cat:** Marketing | Sentence completion and tasks where respondents were ask to project as someone who was brand addicted. Word association/ordering of phrases related to brand addiction projected to someone who was brand addicted. Combined with focus groups. | Qualitative analysis and summary of results into 11 themes. | Sample of n = 19, with 10 males and 9 females, was recruited via a focus group. |
| Dahiya et al. (2014) **Type:** Completion **Cat:** Consumer Sci. | Word association task with different types of bean products. Combined with focus group discussions and structured interviews. | Coding and categorization into 10 categories. Chi-squared test on categories and products. Correspondence analysis. | Sample of n =152. Regional cluster sample from city and villages. Balanced by demographic characteristics. |
| da Silva et al. (2021) **Type:** Completion **Cat:** Consumer Sci. | Story completion involving the purchase of fresh natural cheese. Survey included demographic characteristics and perception of risk. | Categorize responses into categories. Chi-squared tests and cellwise chi-squared tests with respect to regions and split by risk perception. | Sample of n = 326 consumers, clustered by region. |
| Doherty & Nelson (2008) **Type:** Completion, Creative **Cat:** Marketing | "Soft" store attributes found using sentence completion with pictures + conversation bubbles. Case study scenarios (e.g., bad service) with users giving quantitative value for commitment and describing feelings about store. | Content analysis of qualitative answers combined with quantitative loyalty measure used to segment respondents into three loyalty groups (switchers, susceptible, and loyals). Further examine response associations for each loyalty group using chi-squared test. | Cluster sampling of n = 152 consumers in Northern Ireland. |
| Eldesouky et al. (2015) **Type:** Association, Completion **Cat:** Consumer Sci. | Give word associations for cheese packages. Sentence completion with pictures for discussions involving cheese packages. | Coding responses into categories for both tasks and summarize in contingency tables. | Convenience sample of n = 203 recruited from student and other personal databases. |
| Elghannam et al. (2018) **Type:** Association, Completion **Cat:** Consumer Sci. | List items that you would buy and not buy from social media channels. Sentence completion exercises regarding social media channels. | Coding responses into categories from sentence completion and combining/coding of foodstuffs for three different cultures. | A convenience/purposive sample of n = 424 consumers, across three countries (Spain, Mexico, Egypt) who were users of social networks. |



| Esmerino et al. (2017) **Type:** Association, Ordering **Cat:** Marketing | Association task of evaluating cards with different brands of yoghurt/fermented milk packages and giving a range of responses including words, phrases, and pictures. The place cards on paper with closest cards representing most similar products and write 3 to 5 words that differentiate stimuli. | Attributes were categorized and then correspondence analysis (CA) was used to relate categories to product. Variance accounted for by CA solution was reported. Chi-squared tests and cell-wise chi-squared tests were used to test association between categories and products. | Two samples of n = 50 recruited via social networks, email, and advertisements for the association and ordering tasks. A sample of n = 26 was collected for the focus group. |
|---|---|---|---|
| Farhat and Chaney (2024) **Type:** Creative **Cat:** Hospitality | Describe worst experience with a destination then draw a trajectory plot of intensity/polarity of feelings over time. | Qualitative clustering of trajectories into three types and coding of interview responses into different types of brand hate. | Snowball sample of n = 45 collected through travel agencies and including respondents with multiple nationalities. |
| Filep and Greenacre (2007) **Type:** Creative **Cat:** Hospitality | Creative essay assignment, describing a perfect day in their study-abroad town. Combined with a quantitative study on travel motivation. | Essays were coded into motivational themes. Qualitative analysis, including quotes from individual research subjects. | A sample of n = 20, ten males and ten females, partaking in a pre-departure workshop for a study abroad experience in Spain. |
| Font and Hindley (2017) **Type:** Creative **Cat:** Hospitality | Subjects visiting two destinations were asked to select two holiday photos and answer open-ended questions on the photos, e.g., how they relate to subjects' life and travel motives. | Qualitative analysis, including quotes from individual research subjects. | A purposive sample of n = 23 respondents traveling to Svalbard and Venice, with 8 directly asked about motivations and 15 indirectly. |
| Gámbaro (2014) **Type:** Association **Cat:** Marketing | Names of cosmetics (e.g., "nourishing cream") were given as stimuli and users were asked to write down words, images, thoughts and feelings associated with the stimuli. | Terms were coded categorically to give a contingency table. A chi-squared test was run on the table. Correspondence analysis was carried out. | A convenience sample of n = 120 females recruited from shopping centers, universities, and other locations in Uruguay. |
| Gamradt (1995) **Type:** Creative **Cat:** Hospitality | Creative exercise, where school students were asked to give opinions of tourists, give advice to tourists, and draw pictures of imagined tourists. | Coding of student work into categories with multiple categories per student. Qualitative analysis with discussion of pictures and quotes from individual subjects. | A sample of n = 365 students selected in a stratified sample from different categories of school. |
| Gómez-Corona et al. (2016) **Type:** Association **Cat:** Consumer Sci. | Give four words related to craft beer and rank in importance. Give polarity of each word from -3 to 3 on a seven point scale. | Categorization of responses into seven categories. Fisher exact test between categories and countries for craft and industrial beer drinkers. Calculation of polarity index of sentiment. | Intercept sampling of males who consumed beer at least once a month in Paris and Mexico with n = 300 (n = 150 in each location) |



| | | Correspondence analysis (CA) of terms. Hierarchical cluster analysis with coordinates from CA. | and sample balanced between craft and industrial beer drinkers. |
|---|---|---|---|
| Grougiou and Pettigrew (2011) **Type:** Association, Completion **Cat:** Marketing | General views on services with association stimuli (e.g., "dissatisfaction", "quality" and completion stimuli, e.g., ("services are"). | Qualitative summaries, including quotes from individual research subjects. | Saturation was used to determine a sample of n = 60 of a diverse senior consumers. |
| Harrigan et al. (2012) **Type:** Completion **Cat:** Marketing | Sentence completion bubbles for different CRM process scenarios. Combine with quantitative survey on views of e-CRM. | Qualitative summaries, including quotes from individual research subjects. Quantitative data are analyzed with PCA and Chronbach's α. | Sample of n = 1445 SMEs in Ireland. |
| Hilverda et al. (2016) **Type:** Association **Cat:** Consumer Sci. | Give associations with organic food (50% of sample) or organic meat and vegetables (50% of sample). Also measure psychological distance measures using a Likert scale survey. In study 2, rate association "centrality" on quantitative scale. | Individual chi-squared tests on equality of groups for each category across organic usage groups and survey prompt. Run MANOVA + paired comparisons on psychological differences. Calculate adjusted frequencies and median splits. Calculate correlations between centrality scores. | Study 1: Sample of n = 160 in the Netherlands from researcher's networks and from food/organic food internet blogs. Sample split into frequent, occasional, and rare organic food purchasers. Study 2: Convenience sample of n = 52 for rating attributes. |
| Hofstede et al. (2007) **Type:** Association **Cat:** Marketing | Associate clippings of celebrities with brands. Associate each celebrity with a dominant personality characteristic (out of 73 possible characteristics). In second task, associate each brand with a job and each job with a personality characteristic. | Personality characteristics are summarized into dimensions (e.g., competence). Correlations are calculated between tasks for each of the different brands using the proportion of category entries. | A sample of n = 16 of potential beer consumers for mood board and n = 100 for the job association assignment. |
| Holder et al. (2023). **Type:** Association **Cat:** Hospitality | Data analysis consisted of structured interviews discussing pictures showing Indigenous tourism experiences. Projective analysis used to supplement this, e.g., asking what characters in the picture are thinking. | Qualitative analysis of results, including quotes from individual research subjects. | An Australian consumer research panel with n = 41 snowball sample. Projective techniques run with n = 25 subset. |
| Hussey and Duncombe (1999) **Type:** Association **Cat:** Marketing | Associate brands with one of 10 animals and 10 cars (from card sets). Use repertory grid technique to determine adjectives for the stimuli. | Summarize defining constructs for each stimuli. Utilize cluster analysis on attribute values for perceptual similarity between the stimuli. Summarize least | A nonprobability sample of n = 51 subjects. |



| | | and most popular values for each stimuli. | |
|---|---|---|---|
| Jain and Roy (2016)<br>**Type:** Association<br>**Cat:** Marketing | Free word association with favorite celebrities. | Coded responses using content analysis and qualitative content analysis. | A sample of n = 67 subjects from a sample of eight focus groups in India. |
| Ji and King (2018)<br>**Type:** Creation<br>**Cat:** Hospitality | Follow ZMET technique. Prepare eight pictures (from any source) related to dining experience. Combine with eight step ZMET process, including creative description, sorting, and metaphor elaboration and construct elicitation tasks. | Qualitative analysis, including laddering and Kelly's Triadic Sorting (KTS) techniques to help elicit mental maps. | A sample of n = 12 Chinese tourists dining in Portuguese restaurants in the territory of Macau. Sample size determined with qualitative analysis of saturation. |
| Jiménez-Barreto et al. (2020)<br>**Type:** Creation<br>**Cat:** Hospitality | Create stories involving sources of online information used when choosing and booking a holiday destination. Choose and explain choices. Combine with quantitative survey. | Categorize responses and calculate Cohen's kappa. Quantitative data analyzed using two sample tests of means and PLS-SEM. | A paid sample of n = 27 Spanish consumers was recruited using Amazon Mechanical Turk. A sample of n = 307 was used for quantitative analysis. |
| Jung (2022)<br>**Type:** Creation<br>**Cat:** Hospitality | Prepare 6-10 images on fears and risks of travel in the pandemic. Go through seven step ZMET process including a creative task to create a short story. | Development of themes and categories, which are then combined in a consensus map. Constructs derived from the consensus map and summary and quotes from individual subjects for each construct. | Purposive sample of n = 13 subjects who had travel cancelled or postponed due to pandemic, with 11 virtual and 2 in person. |
| Kachersky and Lerman (2013)<br>**Type:** Completion<br>**Cat:** Marketing | Sentence completion to the sentence "marketing is.....". Repeat, but with subjects shown hypothetical marketing materials and with a quantitative measure of concern. | Two judges coded responses for sentiment (positive, negative, neutral) and focus (business, consumer, neither). Chi-squared analysis of categories. Quantitative t-test run on concern across projective categories. | A sample of n = 1006 consumers from a U.S. nationwide panel. A representative sample of n = 160 consumers for follow-up survey. |
| Kim and Kim (2007)<br>**Type:** Creation<br>**Cat:** Hospitality | Gather images related to brand equity. Follow-up interviews, utilizing ZMET process for deeper understanding. | ZMET data were coded into aspects of brand equity using academic coders. A follow-up survey with reliability Chronbach α and factor analysis was used to solidify concepts of brand equity. | Sample of n = 20 hotel managers. Follow up quantitative survey had sample size of n = 350. |
| Kim et al. (2016)<br>**Type:** Creation<br>**Cat:** Marketing | Respondents shown two adverts for luxury brands and were asked to create a story about the people in the brands. Asked questions about the story and asked to have people complete transportation scale and "fashion involvement" scales for participants. | Thematic content analysis to identify themes in the data. Qualitative analysis, including quotes from individual research subjects. | A sample of n = 270 women from Australia, France, and South Korea. Sample tested for similarity of age groups. |
| Lin and Yeh (2023)<br>**Type:** Association, Creative<br>**Cat:** Hospitality | Subjects given VR tourism experience. Asked to choose relevant pictures associated with experience and answer detailed questions on the picture and association with VR tourism as part of ZMET process. | Responses were categorized and organized to create mind maps for respondents. Consensus maps created after analyzing overlaps between metrics. Use quantitative measures of association from subjects to evaluate consensus. | Purposive sampling of people with VR experience and snowball sampling of additional subjects, giving n = 26 subjects. |
| Little and Singh (2014)<br>**Type:** Creation<br>**Cat:** Marketing | Use TET with three scenario pictures on use of the Spanish language in the USA. For each picture scenario the subjects create a story. | Coding of stories, e.g., showing animosity vs. not showing animosity. Chi-squared tests to test differing animosity for the three scenarios and for comparing animosity with demographic characteristics. | Sample of Anglo-Americans from the Midwest of the USA, with n = 54. |
| Ma et al. (2023)<br>**Type:** Association, Creation<br>**Cat:** Hospitality | Follow ZMET technique. Prepare six to ten pictures related to dining experience with service robots. Combine with eight step ZMET process, including creative description, sorting, metaphor elaboration, and construct elicitation tasks. | Coded constructs found in analysis and put constructs and attributes together into conceptual map. | A sample of n = 34 who had dined at a restaurant with service robots or who had heard of service robots |
| Magnini (2010)<br>**Type:** Completion<br>**Cat:** Hospitality | After a virtual tour of a restaurant, complete the sentence "This restaurant is a good place to go when.." | Responses coded as "individualistic" or "collectivist" and a chi-squared test was run for association with the two sample subgroups. | Sample of n =68 from American university, with n = 33 Korean Americans and n = 35 were U. S. born non-Koreans. |
| Mauri (2020)<br>**Type:** Creative<br>**Cat:** Hospitality | Participants asked to collect material including images, adverts, and text on the topic of sustainability. Combine with seven step ZMET process, including | Qualitative analysis of data and coding into concepts. | Recruited sample of n = 10 subjects. |



| | creative description, sorting, and construct elicitation tasks | | |
|---|---|---|---|
| Meliou and Maroudas (2011) **Type:** Association **Cat:** Hospitality | Free association with thoughts and images for the prompt of "tourism development". | Data were simplified and categorized into thematic categories. A Chi-squared test tested association with sample groups. Correspondence analysis of contingency table. | Four groups sampled from students on a postgraduate course (G1 if work experience, G2 otherwise), hospitality workers (G3 if educated, G4 otherwise). Size not stated. |
| Miklavec et al. (2016) **Type:** Association **Cat:** Consumer Sci. | Free word association with three food consumer health/protection symbols. Combine with quantitative conjoint design survey for food symbols. | Categorize responses into appearance vs. meaning and then categorize meaning responses into four different categories. Run chi-squared test of response categories vs. demographic categories. Run ANOVA and conjoint analysis on quantitative data. | Sample of n = 1050 Slovenian adults, with n = 1026 usable word association responses, recruited from a consumer panel and from Facebook |
| Mohamed et al. (2024). **Type:** Association **Cat:** Hospitality | Free word associations with Egyptian food and sentence completion scenarios for aspects (e.g., taste, shape). Subjects shown pictures of Egyptian food and asked which best represented Egyptian food. | Categorization of results into dimensions and attribute dimensions. Creation of radar chart comparing visitors/non-visitors. Creation of a perceptual map (method undisclosed) relating dimensions to visitors/non-visitors. | Combine convenience and snowball sampling to get sample in US Midwest (n = 54) with non-visitors and visitors to Egypt. |
| Mulvey and Kavalam (2010) **Type:** Creation, Ordering **Cat:** Marketing | Collect pictures that represent college experience and then were interviewed on the selection. Order images into piles and gave rationales, along with positive/negative sentiment, for the piles. | Use of the repertory grid technique for further understanding similarities and differences in the experiences. Coding of themes and combining results with results from a laddering analysis. | A sample of n = 50 students from Douglass College, a women's residential college in New Jersey, USA. |
| Mzahi (2014) **Type:** Creation **Cat:** Marketing | Adaption of the "village test". Build a model village and place their house and also the house of the worst enemy. Place service brands in the village to test brand attachment. | Qualitative analysis. There was coding of both attachment to place and attachment to brand. | A sample of n = 21 participants. |
| Oertzen et al. (2020). **Type:** Creation **Cat:** Marketing | Given demographics, "characteristics, likes, dislikes", and "skills, knowledge, abilities" for a hypothetical person, create two personas, one who would be | Analyze and group responses with respect to demographics to create characteristic personas. Combine with qualitative results from an interactive | Purposive sample of n = 38 design professionals for projective techniques and interactive workshop. Combined |



| | interested in co-creation and another who would not. | workshop and PLS-SEM results for quantitative data. | with sample of n = 633 for quantitative work. |
|---|---|---|---|
| Pacheo et al. (2018)<br>**Type:** Association<br>**Cat:** Consumer Sci. | Word association with the prompt "bottled mineral water" combined with quantitative Likert scale survey of views on mineral water + demographic questions. | Words were coded into twelve categories and a chi-squared test and cell-wise chi-squared tests were used to test differing proportions in the categories. PLS-discriminant analysis was used to explain differences between gender. Independent samples t-test on Likert scale data, comparing by gender. | A convenience sample of n = 100 consumers in Brazil. |
| Parente et al. (2023)<br>**Type:** Completion<br>**Cat:** Consumer Sci. | Elicit views of cosmetic product using picture stimuli and story, sentence, and dialog completion scenarios. | Responses to each type of completion scenario were coded and categorized. Categorization was done for each completion scenario. Use chi-squared tests to compare with purchase and demographic categories. | A sample of n = 334 women sampled through social networks. |
| Pich et al. (2015)<br>**Type:** Association, Creative, Completion<br>**Cat:** Marketing | Examine UK political party brands (Conservative/Labour). Word association, picture association, drawing a picture of the UK under each of the parties. Complete sentence bubbles with prompts. | Qualitative discussion and analysis. | A sample of n = 46 British citizens, aged 18-24 years old, and selected from three locations in England.. |
| Pinto et al. (2018)<br>**Type:** Association, Completion<br>**Cat:** Consumer Sci. | Given different shopping lists, each including standard products + type of milk, give buyers' characteristics. Also, word association with stimuli describing milk formulations + quantitative willingness to pay measure. | Coding of categories. Chi-squared between categories and products for both tests. Cell chi-square test on individual measures. ANOVA on quantitative willingness to pay measure. | A convenience sample of n = 450 Brazilian consumers for the shopping list task and n = 100 for the word association task. |
| Pontual et al. (2017)<br>**Type:** Association<br>**Cat:** Consumer Sci. | Free word association for four different types of pizza stimuli. | Chi-squared test of perceived differences in stimuli. Categorization of responses into seven categories. Correspondence analysis (CA) for celiac and non-celiac individuals. Regression vector coefficient to compare CA maps and multiple factor analysis on agreement of celiac/non celiac groups. | A convenience sample of n = 150 people with both celiac and non-celiac individuals. |
| Prayag (2007)<br>**Type:** Association, Creative<br>**Cat:** Hospitality | Word association for South Africa and Cape Town brands. Describe feelings and emotions in "brand fingerprint". Associate colors and people (gender, ethnic group, and age) to the destinations. | Qualitative analysis and summaries of the brands. | A convenience sample of n = 85 international tourists to South Africa who have stayed the minimum of a week. |
| Prebensen (2007)<br>**Type:** Creative<br>**Cat:** Hospitality | Picture and word association exercises. Give associations for pictures of tourism scenes in Northern Norway and word terms related to Northern Norway. Creation of collages. | Qualitative analysis of responses and comparison across country groups. | Sample of n = 38 tourists in France, from France, Germany, Sweden, and Norway |
| Ragelienė (2021)<br>**Type:** Creative<br>**Cat:** Consumer Sci. | Children were asked to draw their favorite lunch or dinner plate (given outline of the plate). Combined with quantitative food preference surveys. | Drawings coded by the size and number of different products (fruit, vegetables, meat, etc.). Independent samples tests for differences in mean values between countries and correlations between quantity and size of different food items. | Sample of n = 480 children from two countries (Denmark and Lithuania) recruited from public school who had completed a parental consent form. |
| Rebollar et al. (2019)<br>**Type:** Association<br>**Cat:** Consumer Sci. | Given stimuli of yogurts with different sweetness characteristics, write down three associations (words, thoughts, feelings). Used as pretest for quantitative survey. | Responses were categorized by three raters and categorizations were evaluated by inter-rater agreement. Results were shown from Biplot (possibly via Correspondence analysis). Compared with multidimensional scaling from quantitative survey. | Convenience sample of n = 112 people. |
| Rocha et al. (2018)<br>**Type:** Association, Completion<br>**Cat:** Consumer Sci. | Word association with six brands of Frankfurter sausage descriptions, along with quantitative survey of preference, feature familiarity, and acceptance of the product description.. | Use emotion corpus to perform basic sentiment analysis, categorize responses and perform chi-squared test and correspondence analysis on category × brand contingency table. Cochran's Q was used for emotional response for testing category emotions across brands and multiple factor analysis (MFA) was used to relate associations with quantitative survey results. Cluster analysis was used on quantitative | Sample in lab respondents of n = 120, with requirement of eating Frankfurter sausages at least once every fifteen days. |



| | | questions. ANOVA test on acceptance results. | |
|---|---|---|---|
| Rojas-Rivas et al. (2020)<br>**Type:** Association<br>**Cat:** Consumer Sci. | Word association with "gastronomy" and listing all foods and ingredients considered part of Mexican cuisine. Combined with semi-structured interviews and demographic information. | Coding of categories. Calculation of cognitive salience index. Using contingency tables of categories vs. demographical variables, chi-squared tests, individual cell chi-squared tests, and correspondence analysis. | Convenience sampling of 22 chefs and 329 consumers, giving a total n = 351 sample. |
| Sass et a. (2018)<br>**Type:** Completion<br>**Cat:** Consumer Sci. | Completion task using picture stimuli and completion for views on different types of egg (e.g., free-range, vitamin enriched). Also ask willingness to pay. | Coding of responses into nine categories, chi-squared test of categories vs. egg type and chi-squared test of individual measures. correspondence analysis of contingency tables. | A sample of n = 100 from Brazil, recruited using a range of methods, including posters, emails, and social networks. |
| Schlinkert et al. (2020)<br>**Type:** Creative, Association<br>**Cat:** Consumer Sci. | Describe an ideal snack and give word associations with healthy and unhealthy snacks | Coding of features and combing words to give simplified document × feature representation. Latent class analysis of ideal snacks and correspondence analysis of word features vs. (healthy, unhealthy, and ideal) snacks. | Sample of n = 1087, convenience sample at consumer "wellness fair" and other places. |
| Soares et al. (2017)<br>**Type:** Association<br>**Cat:** Consumer Sci. | Free word association with "coalho cheese" | Coded into categories using inductive learning. Chi-squared test comparing categories across regions and individual chi-squared tests. | A convenience sample of n = 400 consumers, with 200 taken from each of 2 regions in Brazil. |
| Soulard et al. (2021)<br>**Type:** Creative<br>**Cat:** Hospitality | Draw before and after drawings of personal feelings regarding a transformative tourism experience. Fill in a sentence bubble of thoughts for each experience. Combine with structured interviews. | Qualitative analysis and coding of symbols and associated narratives. | Study recruitment messages sent via tourism organizations, resulting in a final sample of n = 35 subjects. |
| Stach et al. (2017)<br>**Type:** Creative<br>**Cat:** Marketing | Participants were split into three usage groups. Write short stories describing memories of Nutella and then interviewed to elicit additional details. | Qualitative descriptive analysis, contrasting the memories from the three usage groups (non, light, heavy). | A sample of n= 30 German consumers who were no more than 53 years of age and had previously consumed Nutella. Ten each of light, medium, and heavy usage. |
| Tussyadiah and Wang (2016) **Type:** Association, Completion **Cat:** Hospitality | Association tasks comparing smartphone to person and body part, and sentence bubbles with cartoons asking for "think", "feel", and "say" responses to cellphone recommendations. | Coding and categorization of responses into aspects of agency (e.g., intelligence, autonomy), reporting Krippendorff's alpha. Calculation of cooccurrence of associations across types. Summarize responses by category, and look at linkages between categories and stimuli. | A sample of n = 275 students from Hong Kong SAR, China. |
| Ulrich and Tissier-Desbordes (2018). **Type:** Association **Cat:** Marketing | Share words, expressions, and ideas of being a man and being a woman. Combined with discussion of brand elements for masculine and feminine brands and in-depth intervals about respondents life and consumption practices. | Qualitative analysis, including quotes from individual research subjects. | A snowball sample of n = 20, starting with friends and acquaintances. |
| Viana et al. (2014) **Type:** Association, Completion **Cat:** Consumer Sci. | For four different types of burgers, provide four associated words or terms. Likert scale attitudinal questionnaire. | The association terms were coded into 19 groups. Summary of data across all products and for three consumer clusters (found from quantitative questions). | A sample of n = 56 respondents recruited from a Brazilian university. |
| Vidal et al. (2013) **Type:** Association, Completion **Cat:** Consumer Sci. | Free association of words, phrases, and images for the phrase "ready-to-eat salad". A shopping list test, with ready-to-eat salad replaced by base ingredients on second list. A range of picture stimuli and sentence completion bubbles. | Results from all three tests were coded into categories. A chi-squared test was used to compare category counts for two shopping lists. Individual cell chi-squared tests were also implemented. | Subjects recruited from a consumer database held by a Uruguayan university. Convenience sample of n = 169 subjects for word association and n = 151 for sentence completion. |
| Wagner and Peters (2009) **Type:** Creative **Cat:** Hospitality | Create a collage from a pre-determined selection of background colors, pictures, brand logos, slogans, and drawings and illustrations. Asked structured interview questions about destination emotions and branding. | Tabular summary of results by chosen stimuli. Qualitative discussion. | Sample of n = 15 stakeholders in tourism marketing for two Austria tourism destinations. |
| Wassler and Hung (2015) **Type:** Association, Completion **Cat:** Hospitality | Associate country (brand-as-person) and visitors to country (brand-as-user) for USA and Japan, using single word association and sentence completion. | Summary of results in tables and qualitative analysis. | A sample of n = 80 Hong Kong students. |
| Yam et al. (2017)<br>**Type:** Association, Creation<br>Cat: Marketing | Respondents used m-games at least twice and were asked to pick emoticons/emojis to "describe their experience" from a selection. Put emoticons/emojis next to screenshots from game to form collage. Combine with structured interviews. | Qualitative analysis and coding of themes, along with quotes from individual research subjects. | A sample of n = 17 from different households selected to give a diversity of sample. |
| Yoo et al. (2022)<br>**Type:** Creation<br>Cat: Tourism | Respondents asked to prepare pictures on reasons for undertaking a pilgrimage. Retrieve 10-32 pictures per respondent. Work through eight step ZMET process, including selecting most relevant image and opposite image. | Create individual mental maps and an overall consensus mental map. Develop four overall themes from the consensus maps. | Sample Korean protestant Christian adults who have been on a pilgrimage within the last two years. Sample n = 13 from a tour group with 35 people. |
| Youness et al. (2024)<br>Type: Association, Creative<br>Cat: Marketing | Split sample into two groups, prime one affective questions and second cognitive questions. Associate five keywords and choose images after typing keywords into Google image search. Combine into single album and then rate all images/keywords on scale of 1-6. | Qualitative summary and INDSCAL (individual differences scaling) of preference ratings and cluster analysis to create segmented customer maps. | A sample of n = 12 residents of France and who have knowledge and experience of online platforms and who responded to an advertisement. |